\documentclass[prl,twocolumn,preprintnumbers,superscriptaddress]{revtex4-1}
\usepackage{amssymb}
\usepackage{amsmath}
\usepackage{amsfonts}
\usepackage{graphicx}
\usepackage{dcolumn}
\usepackage{bm}
\usepackage{verbatim}
\usepackage{color}
\usepackage{wasysym}
\usepackage{xcolor}
\usepackage{ulem}

\begin{document}

\title{Formation and Dissociation of Field-Linked Tetramers}
\author{Fulin Deng}
\affiliation{School of Physics and Technology, Wuhan University, Wuhan, Hubei 430072, China}
\affiliation{CAS Key Laboratory of Theoretical Physics, Institute of Theoretical Physics, Chinese Academy of Sciences, Beijing 100190, China}

\author{Xing-Yan Chen}
\affiliation{Max-Planck-Institut f\"ur Quantenoptik, 85748 Garching, Germany}
\affiliation{Munich Center for Quantum Science and Technology, 80799 M\"unchen, Germany}

\author{Xin-Yu Luo}
\affiliation{Max-Planck-Institut f\"ur Quantenoptik, 85748 Garching, Germany}
\affiliation{Munich Center for Quantum Science and Technology, 80799 M\"unchen, Germany}

\author{Wenxian Zhang}
\affiliation{School of Physics and Technology, Wuhan University, Wuhan, Hubei 430072, China}
\affiliation{Wuhan Institute of Quantum Technology, Wuhan, Hubei 430206, China}

\author{Su Yi}
\email{syi@itp.ac.cn}
\affiliation{CAS Key Laboratory of Theoretical Physics, Institute of Theoretical Physics, Chinese Academy of Sciences, Beijing 100190, China}
\affiliation{CAS Center for Excellence in Topological Quantum Computation \& School of Physical Sciences, University of Chinese Academy of Sciences, Beijing 100049, China}
\affiliation{Peng Huanwu Collaborative Center for Research and Education, Beihang University, Beijing 100191, China}

\author{Tao Shi}
\email{tshi@itp.ac.cn}
\affiliation{CAS Key Laboratory of Theoretical Physics, Institute of Theoretical Physics, Chinese Academy of Sciences, Beijing 100190, China}
\affiliation{CAS Center for Excellence in Topological Quantum Computation \& School of Physical Sciences, University of Chinese Academy of Sciences, Beijing 100049, China}

\date{\today }

\begin{abstract}
We investigate the static and dynamic properties of tetratomic molecules formed by two microwave-shielded polar molecules across field-linked resonances. In particular, we focus on two-body physics and experimental techniques unexplored in the recent experiment [X.-Y. Chen {\it et al}., Nature {\bf626}, 283 (2024)]. We show that, compared to the lowest tetramer state, higher tetramer states typically have longer lifetimes, which may facilitate a further cooling of tetramer gases towards quantum degeneracy. To detect tetramers, we identify the distinctive time-of-flight images from ramp dissociation, which can be observed by lowering the ramp rate of the microwave. Remarkably, in the modulational dissociation of tetramers, we find that multi-photon processes induce dissociation even below the threshold modulation frequency when the modulation amplitude is sufficiently high. Given the universal form of the inter-molecular potential for microwave-shielded polar molecules, our results also apply to other molecular gases widely explored in recent experiments.
\end{abstract}
\maketitle

\textit{Introduction.}---Ultracold molecules~\cite{Carr2009,Ye2017}, with their abundant internal and external degrees of freedom, offer an exceptional avenue for exploring quantum dynamics~\cite{Koch2019,Liu2023}, controlling chemical reactions~\cite{Krem2008,Ni2019,Liu2022}, and enabling a wide range of applications in quantum computing~\cite{DeMille2002,Cornish2020}, quantum simulation~\cite{Zoller2006,Zwierlein2021}, quantum information processing~\cite{Zoller2006a,Tesch2002,Wall2015,Albert2020}, and precision measurement~\cite{Kozlov2007,Berger2010,Hinds2011,Hutzler2020}. After over a decade of extensive efforts~\cite{Ye2008,Nagerl2014,Cornish2014,Zwierlein2015,Bloch2018,Ospelkaus2020,Wang2016,Jamison2017,Will2022,Ni2021,Tarbutt2021,Ye2020a,Ye2021,Ye2020b,Doyle2021}, ultracold Bose and degenerate Fermi gases of molecules have been successfully achieved in experiments~\cite{Ye2019,Luo2021,Luo2022a,Luo2022b,Cao2022,Wang2023,Will2023,Will2023b}
through microwave shielding techniques. The stable molecular gases provide an ideal platform for investigating novel scattering processes~\cite{Quemener2018,Park2023,Tang2023,Chen2023} and intriguing many-body effects~\cite{Pfau2009,You1999,Baranov2002,Shi2010,Pu2010,Hirsch2010,Shlyapnikov2011,Baranov2012,Shi2014,Zhai2013,Deng2023} induced by long-range anisotropic dipole-dipole interactions (DDI).

Remarkably, both DDI and the short-range shielding potential
between microwave-shielded molecules (MSMs) can be effectively manipulated by the
elliptic angle and Rabi frequency of the microwave field, leading to field-linked (FL) scattering resonances~\cite{Luo2022b,Deng2023}. The adiabatical ramping of microwave field across FL resonances, a tetratomic molecule composed of two MSMs can be formed. Although the tetramer has been proposed 
theoretically for bosonic molecules~\cite{Quemener2023}, the first stable ultracold FL tetramer (NaK)$_2$ is realized experimentally~\cite{Chen2023} in a fermionic NaK gas, benefiting from Pauli-blocking and reduced collisional losses~\cite{Bause2023} via microwave-shielding. The fermionic MSMs emerge as promising candidates for investigating unprecedented strongly correlated physics in the crossover from $p$-wave Bardeen-Cooper-Schrieffer (BCS) superfluidities of molecules to
Bose-Einstein condensations (BEC) of tetramers~\cite{Shi2014,Zhai2013,sademelo1993,sademelo2006}.

In this Letter, we investigate the static and dynamic properties of the FL tetramer formed by two fermionic MSMs, focusing on properties and experimental techniques unexplored in the recent experiment~\cite{Chen2023}.
We study the binding energy and lifetime of $p_x$ tetramers which differ in spatial symmetry from the $p_y$ tetramers realized in~\cite{Chen2023}.
Remarkably, the longer lifetimes of $p_x$ tetramers (e.g., 10.4s for LiRb molecules) show promise for cooling tetramers to quantum degeneracy. By studying tetramer dissociation, we find that the featured time-of-flight images of tetramer wavefunctions can be observed in ramp dissociation (RD)
by lowering the microwave ramping rate. In modulational dissociation (MD), large modulations of the elliptic angle lead to tetramer dissociation below the threshold modulation frequency, linked to multiple-photon processes.

\begin{figure}[tbp]
\includegraphics[trim=0 0 90 16, clip, width=1\columnwidth]{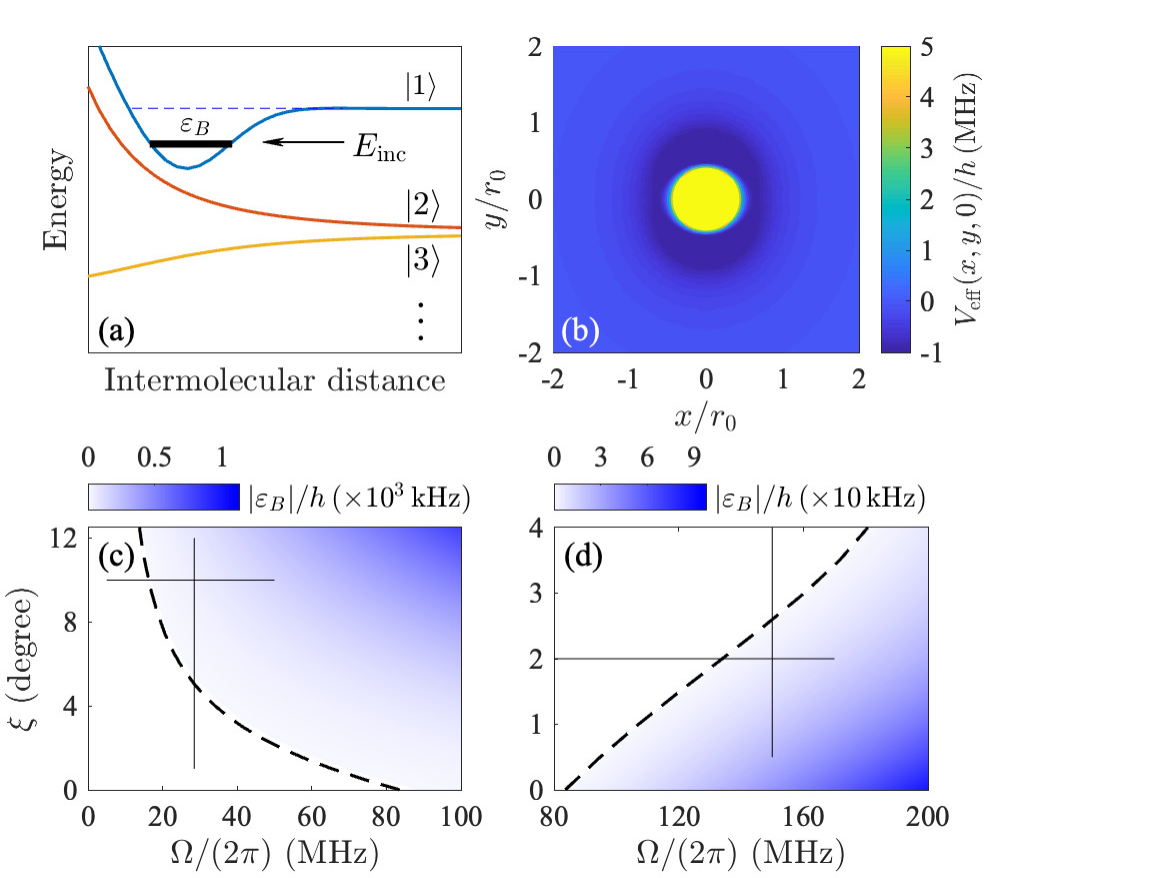}
\caption{(a) Schematic of the adiabatic curves of two interacting molecules. The bold black line denotes a bound state and the arrow marks the incident energy. (b) A typical effective potential on the $x$-$y$ plane with $\Omega/(2\pi)=130 \mathrm{MHz}$ and $\protect\xi=2^{\circ}$. (c) and (d) are distributions of the binding energies for, respectively, the $p_y$- and $p_x$-tetramer states on the $\Omega$-$\xi$ plane. The dashed lines in (c) and (d) denote the positions of shape resonances. The vertical and horizontal lines denote the ranges of the control parameters being ramped in tetramer association and dissociation.}
\label{BS}
\end{figure}

\textit{Models for two MSMs}.---As a concrete example, we consider the NaK molecules which are treated as rigid rotors with electric dipole moment $d$. At ultracold temperature, only the lowest four rotational manifolds that consist of four rotational states are of relevance. To achieve microwave shielding, molecules are illuminated by a microwave field blue detuned from the transition frequency of the lowest and the first excited rotational levels. The interaction potential between molecules in the highest dressed state, $|+\rangle$, is then consistently repulsive at short distance, preventing molecules from forming an unstable complex.

For the model of two molecules, we first note that the inter-molecular DDI is
\begin{align}
V({\mathbf{r}})=\frac{d^{2}}{4\pi \epsilon _{0}r^{3}}\left[ \hat{\mathbf{d}}
_{1}\cdot \hat{\mathbf{d}}_{2}-3(\hat{\mathbf{d}}_{1}\cdot \hat{\mathbf{r}})(
\hat{\mathbf{d}}_{2}\cdot \hat{\mathbf{r}})\right] ,
\end{align}%
where $\epsilon_{0}$ is the electric permittivity of vacuum, $r=|{\mathbf{r}}|$, $\hat{\mathbf{r}}={\mathbf{r}}/r$, and $\hat{\mathbf{d}}_{j=1,2}$ is the unit vector along the internuclear axis of the $j$th molecule. Then in the two-molecule Hilbert space with basis states formed by symmetrized dressed states, the $|++\rangle$ state only couples to six lower two-molecule states through DDI~\cite{Deng2023,SM}. For brevity, we label these seven basis states as $|\nu\rangle$ with $\nu=1$ to $7$ in descending order of energies ($E_\nu^{(\infty)}$) of the asymptotic states. In Fig.~\ref{BS}(a), we schematically show the adiabatic potentials obtained by diagonalizing the interaction matrix in the 7-dimensional Hilbert space. 

Although the complete description of two interacting $|+\rangle$ molecules involves all seven channels, the situation is greatly simplified at ultracold temperatures for which the typical timescale for the rotation is much smaller than that of the center-of-mass motion. Consequently, we may focus on the highest potential curve which is the effective potential between MSMs. As shown by us in Ref.~\cite{Deng2023}, the effective potential can be approximated as
\begin{align}
V_{\mathrm{eff}}(\mathbf{r})&=\frac{C_{6}}{r^{6}}\sin ^{2}\theta \left\{ 1-%
\mathcal{F}_{\xi }^{2}(\varphi )+[1-\mathcal{F}_{\xi }(\varphi )]^{2}\cos
^{2}\theta \right\}  \notag \\
&\quad+\frac{C_{3}}{r^{3}}\left[ 3\cos ^{2}\theta -1+3\mathcal{F}_{\xi
}(\varphi )\sin ^{2}\theta \right] ,  \label{Vps}
\end{align}
where $\mathcal{F}_{\xi }(\varphi )=\sin 2\xi \cos 2\varphi $ with $\xi$ being the ellipticity of the microwave, $\theta$ ($\varphi$) is the polar (azimuthal) angle of ${\mathbf{r}}$, and $C_3=d^{2}\Omega^2/\left[ 48\pi \epsilon _{0}(\Omega^2+\delta^{2})\right]$ with $\Omega$ being the Rabi frequency and $\delta$ the detuning of the microwave~\cite{SM}. As to $C_6$, although it is convenient to use the approximate expression $C_6 \approx 18(\Omega^2+\delta^2)^{1/2}C_3^2$ to analyze the properties of $V_{\rm eff}$, we adopt, in all numerical calculations, a more accurate $C_6$ by fitting the adiabatic potential~\cite{Deng2023}. It should also be noted that Eq.~\eqref{Vps} is valid only when $|\xi|\apprle15^{\circ}$ and $r^{3}>d^{2}/(4\pi \epsilon_{0}\Omega)$. 

Among the control parameters of $V_{\rm eff}$, we shall, for simplicity, fix the detuning at $\delta=-2\pi\times 9.5\,{\rm MHz}$ throughout this work, which reduces the control parameters to $\xi$ and $\Omega$. Without loss of generality, we always assume that $\xi\geq 0$. For $\xi=0$, $V_{\rm eff}({\mathbf r})$ is axially symmetric with a circular attractive potential well on the horizontal planes in the vicinity of $z=0$. However, as shown in Fig.~\ref{BS}(b), this axial symmetry is broken for a nonzero $\xi$ such that the depth of the attractive potential well is increased (lowered) along the $y$ ($x$) direction. Moreover, for a given $\xi$, the depth of the potential well grows as $\Omega$ is increased. These tunabilities, as demonstrated in Refs.~\cite{Deng2023,Chen2023}, can be used to induce scattering resonances and form bound states.

Now the relative motion of two MSMs is described by the Hamiltonian 
\begin{align}
H_{\rm eff}=-\frac{\hbar^2\nabla^2}{M}+V_{\rm eff}({\mathbf r}),\label{heff}
\end{align}
where $M$ is the mass of the molecule. The validity of this single-channel model was justified by studying scatterings of two $|+\rangle$ molecules~\cite{Deng2023}. 

\textit{Properties of the tetramer states}.---The wavefunction $\psi_B$ of a tetramer states can be obtained by numerically solving the Schr\"odinger equation
\begin{equation}
H_{\rm eff}\psi_B=\varepsilon_B\psi_B,    
\end{equation}
 where $\varepsilon_B$ ($<0$) is the binding energy. For $\xi=0$, the $z$ component of the angular momentum is conserved such that the eigenstates of $H_{\rm eff}$ have a definite quantum number $m$ for the projection of the angular momentum. It is numerically found that bound states emerge when $\Omega$ is larger than the critical value $\Omega_c\equiv 2\pi\times83\,\mathrm{MHz}$. These bound states are doubly degenerate with respect to $m=\pm1$ and the bound-state wave functions are $\psi_{B}^{(\pm)}(\mathbf{r})=\sum_{{\rm odd}\;l}Y_{l,\pm 1}(\hat{r})\phi_{l}(r)$, where the radial wave functions $\phi_l$ are determined numerically. For $\xi>0$, the degeneracy is lifted and leads to a $p_y$-tetramer state $\psi_{B}^{(y)}\propto (\psi_{B}^{(+)}+\psi_{B}^{(-)})$ and a $p_x$-tetramer state $\psi_{B}^{(x)}\propto (\psi_{B}^{(+)}-\psi_{B}^{(-)})$ with distinct critical Rabi frequencies $\Omega_c^{(y)}$ and $\Omega_c^{(x)}$, respectively. Because the wave function of the $p_y$ ($p_x$)-tetramer state has a larger amplitude along the $y$ ($x$) direction compared to that along the $x$ ($y$) direction, we have $\Omega_c^{(y)}\leq\Omega_c\leq\Omega_c^{(x)}$. In Figs.~\ref{BS}(c) and (d), we map out $|\varepsilon_{B}|$ in the $\Omega $-$\xi$ parameter plane for $\psi_B^{(y)}$ and $\psi_B^{(x)}$, respectively. The dashed lines denote the position of the shape resonance, i.e., the critical Rabi frequencies. As expected, $\Omega_c^{(y)}$ ($\Omega_c^{(x)}$) decreases (increases) with the increase $\xi$.
The dipole moment of tetramer states is approximately twice that for the $|+\rangle$ molecule~\cite{SM}. 

\begin{figure}[tbp]
\includegraphics[width=1.0\linewidth]{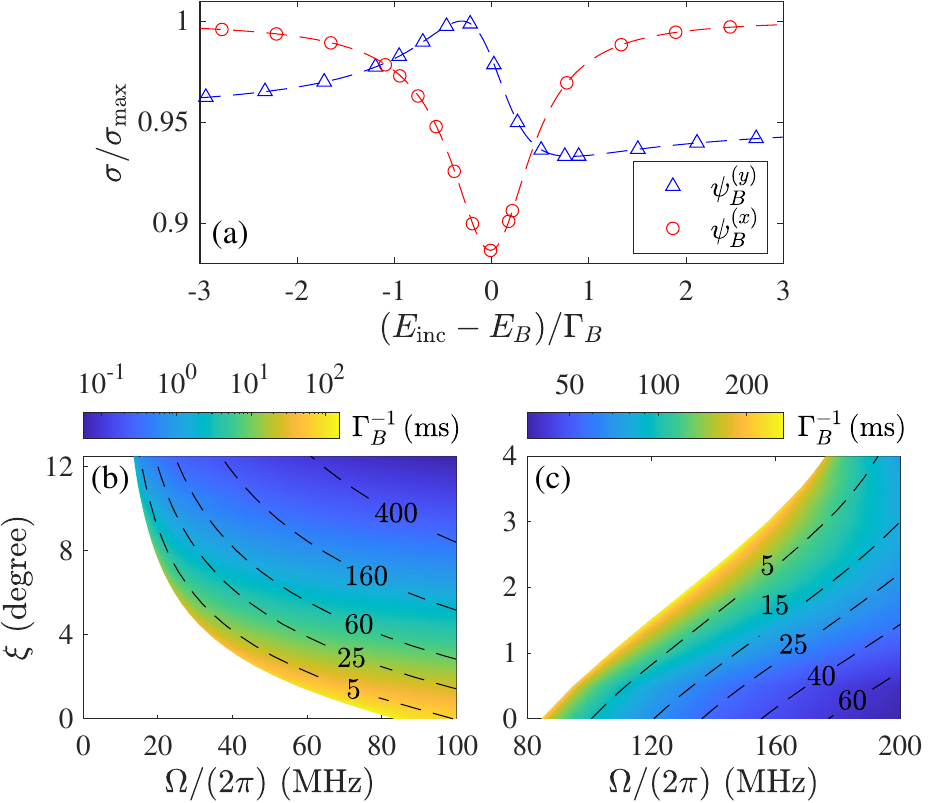}
\caption{(a) Scattering cross sections across the resonances associated with $\psi_B^{(y)}$ (`$\triangle$') and $\psi_{B}^{(x)}$ (`$\ocircle$') with parameters $(\Omega/(2\pi),\xi)=(19.52\ \mathrm{MHz},10^{\circ })$ and $(148.94\ \mathrm{MHz},2^{\circ })$, respectively. Lifetimes $\Gamma_B^{-1}$ (color maps, in units of ${\rm ms}$) and binding energy $|E_B|/h$ (contour lines, in units of ${\rm kHz}$) for the bound states $\psi_B^{(y)}$ (b) and $\psi_B^{(x)}$ (c).}
\label{lineshape}
\end{figure}

Since a tetramer can dissociate into two molecules in the lower six channels, it has a finite lifetime. To determine the lifetimes of the tetramer states, we consider the collisions of two molecules interacting via $V({\mathbf r})$ with incident energy $E_{\rm inc}$ lying between $E_1^{(\infty)}$ and $E_2^{(\infty)}$ [see Fig.~\ref{BS}(a)]. By tuning incident energy, scattering resonances can be induced when $E_{\rm inc}$ is in resonance with binding energies of the tetramer states. The lifetime of the corresponding tetramer state is associated with the width of the resonance. The above scenario is equivalent to a Feshbach resonance with $\nu=1$ and $\nu>1$ being the closed and open channels, respectively. The scattering cross-section in the vicinity of a Feshbach resonance takes the form~\cite{Taylor1972}
\begin{equation}
\sigma (E_{\rm inc})=\frac{2\pi }{k_{2}^{2}}\left\vert ig^{2}G(E_{\rm inc})+S_{\mathrm{bg}%
}-1\right\vert ^{2},  \label{sigma}
\end{equation}
where $k_{2}=\sqrt{2m\left(E_{\rm inc}-E_2^{(\infty)}\right)/\hbar^2}$ is the incident momentum with respect to the channel $|2\rangle$, $g$ is the effective coupling between $|1\rangle $ and $|2\rangle $, and $S_{\mathrm{bg}}$ is the background scattering amplitude of molecules in the channel $|2\rangle $. Finally, $G(E)=1/(E-E_{B}+i\Gamma _{B}/2)$ is the propagator describing a transient process that two colliding molecules form a tetramer with binding energy $E_{B}$ and lifetime $1/\Gamma _{B}$.

To find the scattering cross section, we perform the coupled-channel scattering calculations which involve all $7$ channels~\cite{Deng2023}. For completeness, we include a van der Waals potential $-C_{\mathrm{vdW}}/r^{6}$ in $V({\mathbf{r}})$ to describe the universal background scattering~\cite{Idziaszek2010}. Moreover, to account for the loss caused by the formation of the four-body complex, we place an absorption boundary condition inside the shielding core~\cite{Clary1987}. Figure~\ref{lineshape}(a) plots the typical cross section $\sigma$ for the scattering from channel $(\nu lm)=(210)$ to $(210)$ with parameters in the vicinity of the bound state $\psi_B^{(y)}$ (`$\triangle$') and $\psi_B^{(x)}$ (`$\ocircle$'), respectively. The dashed lines are the corresponding fits to Eq.~\eqref{sigma}, from which one determines the binding energies $E_B$ and decay rate $\Gamma_B$. With $E_B$, we can now calibrate $\varepsilon_B$, i.e., the binding energy obtained via the effective potential. As shown in SM~\cite{SM}, for all control parameters covered by our calculations, the relative difference between $\varepsilon_B$ and $E_B$ is always below $10\%$, which further confirms the validity of the effective potential~\cite{Deng2023}.

In Fig.~\ref{lineshape}(b) and (c), we map out the lifetime $\Gamma_B^{-1}$ on the $\Omega$-$\xi$ plane for bound states $\psi_B^{(y)}$ and $\psi_B^{(x)}$, respectively. As a comparison, we also present the corresponding binding energy $|E_B|$ using contour lines. A general observation is that smaller binding energy leads to a larger lifetime. This phenomenon can be roughly understood as follows: A shallower bound state in the closed channel implies a larger incident momentum (or, equivalently, a faster oscillating scattering wave function) in the open channel, which leads to a smaller Frank-Condon factor between closed and open channels. As a result, the decay rate $\Gamma_B$ becomes smaller. However, a closer look at the contour lines reveals that, even with the same binding energy, the lifetime may vary significantly. For instance, along the contour line with $|E_B|/h=25\,{\rm kHz}$ in Fig.~\ref{lineshape}(b), the lifetime is extended from $0.4\,\mathrm{ms}$ to as large as $24\,{\rm ms}$ via increasing $\Omega$. Here, a larger Rabi frequency gives rise to a larger energy spacing between asymptotic states of open and closed channels, which effectively improves the shielding effect. Another interesting observation is that the $p_x$-tetramer state usually has a longer lifetime than its $p_y$ counterpart. This becomes particularly prominent for molecules with larger dipole moments. For example, the lifetime of the $p_x$-tetramer state of two LiRb molecules can be as long as $10.4\,{\rm s}$ under suitable control parameters~\cite{Dulieu2005,SM}. The sufficiently long lifetime of the tetramers allows one to perform the evaporative cooling of the tetramer gases, which may eventually lead to Bose-Einstein condensations of tetramers. 

\textit{Association and dissociation of tetramers}.---We now turn to study the dynamics of the association and dissociation of the tetramers by using the single-channel model Eq.~\eqref{heff}. In general, both association and dissociation can be realized by tuning either $\xi$ or $\Omega$. For simplicity, we shall only present the results for tuning $\xi$, and the results for varying $\Omega$ can be found in SM~\cite{SM}.

\begin{figure}[tbp]
\includegraphics[width=\linewidth]{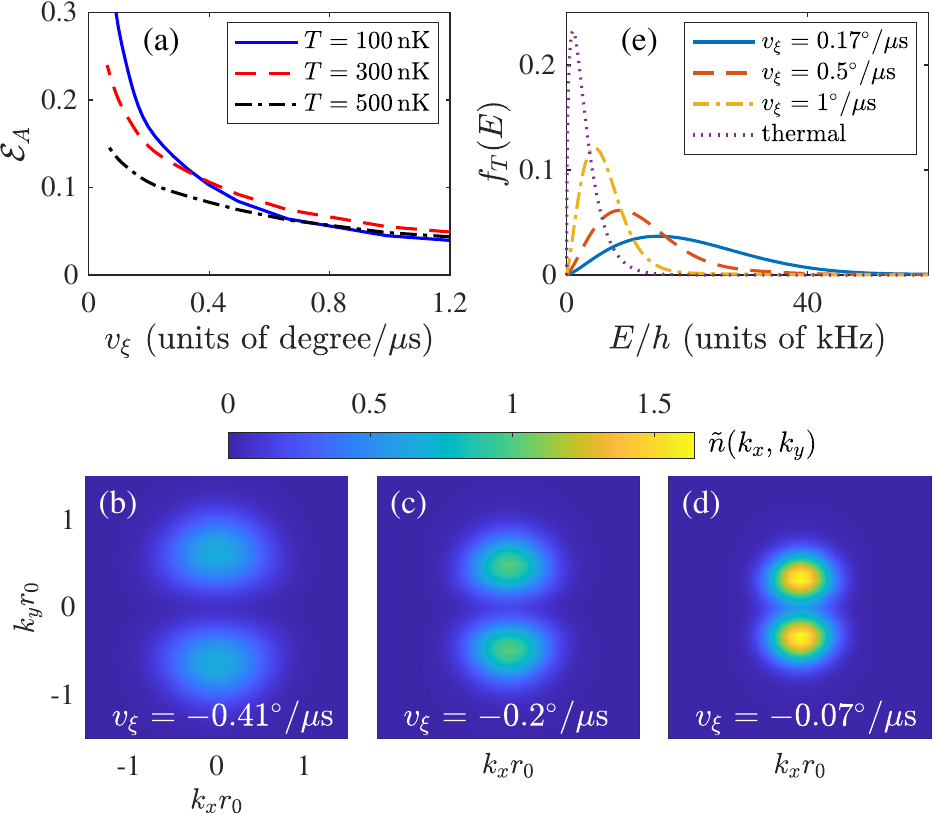}  
\caption{(a) Ramping rate dependence of the association efficiency for various temperatures with $\Omega/(2\pi)= 150\,\mathrm{MHz}$ and $\xi$ is ramped up from $0.5^\circ$ to $5^\circ$. (b)-(d) Integrated momentum distributions, $\tilde n(k_x,k_y)$ (in units of $r_0^{2}$), of the dissociated molecules from a tetramer gas with $T=100{\rm nK}$ for various ramping rates. (e) Energy distributions $f(E)$ corresponding to (b-d). The dotted line represents the energy distribution of a thermal gas at $T=100\,{\rm nK}$. For both association and dissociation, $\xi$ is ramped in the range $[0.5^\circ,5^\circ]$.}
\label{ddiss}
\end{figure}

To associate a tetramer, we start from a two-molecule scattering state $|\psi_{\mathbf k}\rangle$ of relative momentum $\mathbf k$ under initial $\xi$. Then $\xi$ is ramped up with a constant ramp rate $v_\xi\equiv d\xi/dt$ across a shape resonance. The transition probability to the bound state $|\psi_B\rangle$ corresponding to the final $\xi$ is 
\begin{align}
p_{a}(\mathbf{k})=\left\vert \left\langle \psi_{B}\right\vert {U(\tau)} \left\vert \psi _{\mathbf{k}}\right\rangle \right\vert ^{2}, 
\end{align}
where $\tau$ is ramp time and $U(\tau)=\mathcal{T}\exp [-i\int_{0}^{\tau}dtH_{\mathrm{eff}}(t)]$ is the time-evolution operator with $\mathcal{T}$ denoting the time-ordered exponential. Now, for a molecular gas with density $n$ and temperature $T$, the \textit{association efficiency} to tetramers can be numerically evaluated using~\cite{Burnett2007,Greene2019}
\begin{equation}
\mathcal{E}_{A}=2n\lambda_{T}^{3}\int d\mathbf{k}e^{-\beta k^{2}/M}p_{a}(\mathbf{k}),
\end{equation}
where $\beta=1/(k_BT)$ is the inverse temperature with $k_B$ being the Boltzmann constant and $\lambda _{T}=\sqrt{4\pi \hbar^2 /(M k_{B} T)}$ is the thermal de Broglie wavelength.

In Fig.~\ref{ddiss}(a), we plot the $v_\xi$ dependence of $\mathcal{E}_{A}$ for molecular gases under various temperatures. The association efficiency is a monotonically decreasing function of the ramping rate, indicating that adiabaticity is crucial for achieving higher productivity of tetramers. Interestingly, for the temperature dependence of the conversion efficiency, it is found that $\mathcal{E}_A$ is not a monotonically decreasing function of $T$, which is somehow in contrast with the intuition since $\lambda_T$ is a decreasing function of $T$. A rough explanation is that the thermal distribution should also match $p_a(\mathbf{k})$ to obtain higher $\mathcal{E}_A$. As a result, $\mathcal{E}_A$ does not necessarily increase as $T$ decreases.

The experimental detection of tetramers can be carried out by first dissociating them into molecules using either RD or MD. For RD, $\xi$ is ramped down to the scattering-state regime with ramping rate $v_\xi$ in a time interval $\tau$. The probability of finding a molecule pair with relative momentum $\mathbf{k}$ is then 
\begin{align}
p_{\rm RD}(\mathbf{k})=\left\vert \langle \mathbf{k}|U(\tau)|\psi_{B}\rangle\right\vert^{2},
\end{align}
where $|\psi_B\rangle$ is the bound state corresponding to the initial $\xi$ and $U(\tau)$ is now the time-evolution operator for the dissociation dynamics. Then, for a tetramer gas at equilibrium with temperature $T$, the momentum distribution of the dissociated molecules is
\begin{align}
n_T(\mathbf{k})=\int d\mathbf{p}\frac{e^{-\beta(\mathbf{k}+\mathbf{p})^{2}/(4M)}}{(4\pi M k_{B} T)^{3/2}}p_{\rm RD}\left(\frac{\mathbf{k}-\mathbf{p}}{2}\right),\label{temptetr}
\end{align}
which is experimentally detectable via the time-of-flight (TOF) imaging. In Figs.~\ref{ddiss}(b-c), we present the integrated momentum distribution, $\tilde n_T(k_x,k_y)=\int dk_zn_T({\mathbf k})$, dissociated from a tetramer gas with $T=100\,{\rm nK}$ at various ramping rates. As can be seen, all momentum distributions are featured by the double peaks, in striking contrast to the narrow crescent shape in the TOF images of MD~\cite{Chen2023}. Moreover, because the smaller the ramping rate, the higher the visibility, it is more feasible in the experimental detection by adopting smaller ramping rates. We may also distinguish the dissociated molecules by examining their energy distributions, i.e., $f_T(E)=\int d\mathbf{k}\delta (E-k^{2}/(2M))n_T(\mathbf{k})$. In Fig.~\ref{ddiss}(e), we plot the energy distributions of the dissociated molecules corresponding to different ramping rates. As a comparison, we also plot the thermal distribution with $T=100\,{\rm nK}$. Compared to the thermal gas, $f_T(E)$ of the dissociated molecules is significantly broadened, which can be used as the signature of the tetramer even when the double-peak structure is not observed.
\begin{figure}[tbp]
\includegraphics[width=\linewidth]{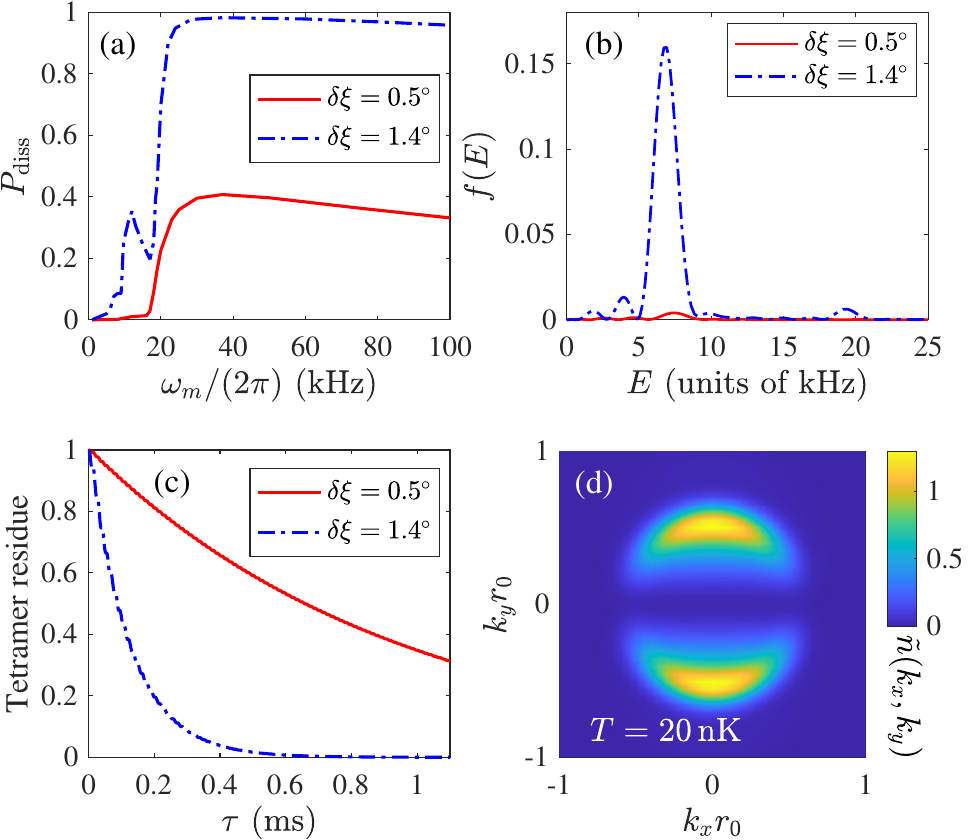}
\caption{MD of a $p_y$ tetramer prepared at $\Omega=(2\pi)\times 28.5\,\mathrm{MHz}$ and $\xi_{0}=8^{\circ}$. The binding energy of the initial state is $|\varepsilon_B|/h=17\,{\rm kHz}$. (a) $P_{\rm diss}$ versus $\omega_m$ for $\tau=0.5\,\mathrm{ms}$ and different $\delta\xi$'s. (b) Energy distributions of a pair of dissociated molecules with $\omega_m=2\pi\times 12.5\,{\rm kHz}$, $\tau=0.5\,{\rm ms}$, and various $\delta\xi$'s. (c) $1-P_{\rm diss}$ versus $\tau$ at $\omega_m=2\pi\times 37\,{\rm kHz}$ for two $\delta\xi$'s. (d) Integrated momentum distributions for $\omega_m=2\pi\times 37\,{\rm kHz}$, $\delta\xi=1.4^\circ$, $\tau=0.125\,{\rm ms}$, and  $T=20\,{\rm nK}$.} 
\label{mdiss}
\end{figure}

Observations of tetramers are experimentally realized through MD, for which the elliptic angle is periodically modulated around $\xi_0$, i.e., ${\xi}(t)=\xi_{0}+\delta \xi \sin \omega_{m}t$, where $\delta \xi $ and $\omega_{m}$ are the amplitude and frequency of the modulation, respectively. The underlying mechanism of MD is in analogy to the dissociation of Feshabch molecules by modulating the magnetic field~\cite{dFeshbach1,dFeshbach2}. Here we present a detailed analysis. A MD starts with a bound state $|\psi_B\rangle$ prepared under $\xi_0$. The \textit{dissociation probability} after being modulated for a driving time $\tau$ is $P_{\rm diss}=\langle\psi(\tau)|\mathcal{P}_{\rm sc}|\psi(\tau)\rangle$, where $|\psi(\tau)\rangle=U(\tau)|\psi_B\rangle$ with $U(\tau)$ being the time-evolution for MD and $\mathcal{P}_{\rm sc}=1-|\psi_B\rangle\langle\psi_B|$ is a projection operator. The dissociation probability can then be numerically calculated straightforwardly. Interestingly, the dynamics of the system under a small amplitude modulation can be captured by a time-independent Hamiltonian~\cite{SM}.

Figure~\ref{mdiss}(a) plots the dissociation spectra of a $p_y$-tetramer state for various $\delta\xi$'s. The binding energy of the initial tetramer is $|\varepsilon_B|/h\simeq 17\,\mathrm{kHz}$. For $\delta\xi=0.5^\circ$, $P_{\rm diss}$ first slowly grows with $\omega_m$ when $\omega_m$ is smaller than the threshold value $\omega_m^*=|\varepsilon_B|/\hbar$. In this regime, the dissociation is mainly due to the off-resonant coupling between the bound state and scattering states. For $\omega_m>\omega_m^*$, the tetramer is resonant to scattering states such that $P_{\rm diss}$ quickly increases and reaches a peak value. Then for $\delta\xi=1.4^\circ$, although the dissociation spectrum above the threshold only quantitatively differs from that of $\delta\xi=0.5^\circ$, the below-threshold behavior is completely different. In fact, a dissociation peak emerges at $\omega_m=2\pi\times 12.5\,{\rm kHz}$, which, as shown below, is due to the multiple-photon excitation under a strong modulation. To show this, let us examine the energy distribution of the dissociated molecules, i.e., $f_0(E)=\int d\mathbf{k}\delta (E-k^{2}/M)p_{\rm MD}(\mathbf{k})$, where $p_{\rm MD}({\mathbf k})=|\langle{\mathbf k}|\mathcal{P}_{\rm sc}|\psi(\tau)\rangle|^2$ is the momentum distribution of two dissociated molecules. Figure~\ref{mdiss}(b) plots $f_0(E)$ for $\omega_m=2\pi\times 12.5\,{\rm kHz}$ and different $\delta\xi$'s. For $\delta\xi=1.4^{\circ}$ a prominent peak at $E/h\approx 2\times12.5-17=8\,{\rm kHz}$ is observed, which indicates the dissociation of the tetramer to two molecules after absorbing two photons. Eventually, the two-photon process gives rise to a below-threshold peak on the dissociation spectrum.

Furthermore, we plot, in Fig.~\ref{mdiss}(c), dissociation probability as a function of driven time for $\omega_m=2\pi\times 37\,{\rm kHz}$ and with two distinct $\delta\xi$'s. In both cases, the dissociation probability decay exponentially as $P_{\rm diss}=1-e^{-\gamma_{\rm diss}\tau}$, where the dissociation rates are $\gamma_{\mathrm{diss}}=1.06$ and $8.27\,\mathrm{ms}^{-1}$ for $\delta\xi=0.5^\circ$ and $1.4^\circ$, respectively. These results confirm that the above-threshold MD is essentially a spontaneous emission for which the Markovian approximation is valid~\cite{SM}. Finally, for experimental detection, we consider the momentum distribution of the dissociated molecules from a tetramer gas of temperature $T$, which can be obtained by replacing $p_{\rm RD}({\mathbf k})$ in Eq.~\eqref{temptetr} by $p_{\rm MD}({\mathbf k})$. In Fig.~\ref{mdiss}(d), we plot the integrated momentum distribution $\tilde n_T(k_x,k_y)$ with $T=20\,{\rm nK}$. As can be seen, the momentum distribution is featured by the crescent shape, in agreement with the experimental observations. 

\textit{Conclusion}.---We have studied the static and dynamic properties of FL tetramers formed by two fermionic MSMs using the effective potential. Our studies reveal that the higher $p_x$-tetramer states generally exhibit longer lifetimes than the $p_y$-tetramer states, potentially facilitating further cooling of tetramer gases to quantum degeneracy. Additionally, we propose a novel method for the experimental detection of tetramers by using RD. In the case of MD with large amplitude modulations, we have identified below-threshold peaks in the dissociation spectrum, which are linked to multiple-photon processes. Given the universal form of the inter-molecular potential, our findings are directly applicable to other fermionic molecules, significantly enriching the two- and many-body physics of ultracold molecular gases. Our work opens promising avenues for realizing tetratomic BEC through evaporative cooling and exploring the diverse many-body physics, including the fascinating BCS-BEC crossover in MSMs. 

\textit{Acknowledgment}.---X.-Y. C. and X.-Y. L. thank the MPQ \textit{NaK molecules} team
for fruitful discussions. This work was supported by the NSFC
(Grants No. 12135018, No.12047503, and No. 12274331), by National Key Research
and Development Program of China (Grant No. 2021YFA0718304), and by CAS Project for Young Scientists in Basic Research (Grant No. YSBR-057).
X.-Y. C. and X.-Y. L. acknowledge support from the Max Planck Society, the
European Union (PASQuanS Grant No. 817482) and the Deutsche
Forschungsgemeinschaft under German Excellence Strategy -- EXC-2111 --
390814868 and under Grant No.\ FOR 2247.

\bibliography{ref_dMolecule1.bib}

\clearpage

\widetext

\begin{center}
\textbf{\large Supplemental Materials: }
\end{center}

This Supplemental Material is structured as follows. In the first section,
we derive the effective potential $V_{\mathrm{eff}}$ for two molecules in
the highest dressed state channel and dipole moments of tetramers. In the second section, we employ the
log-derivative method to calculate scattering amplitudes, cross sections, the binding energies and the life-time of tetramers
using the 7 channel model, where the validity
of $V_{\mathrm{eff}}$ is justified. In the third section, we calculate the association efficiency and the momentum (energy) distribution in the ramp dissociation. In the fourth section, we derive the effective Hamiltonian describing the modulational dissociation. In the fifth section, we calculate the dissociation spectrum and the momentum distribution in the modulational dissociation.

\setcounter{equation}{0} \setcounter{figure}{0} \setcounter{table}{0} %
\setcounter{page}{1} \setcounter{section}{0} \makeatletter%
\renewcommand{\theequation}{S\arabic{equation}} \renewcommand{\thefigure}{S%
\arabic{figure}} \renewcommand{\bibnumfmt}[1]{[S#1]} \renewcommand{%
\citenumfont}[1]{S#1} \renewcommand{\thesection}{S\arabic{section}}%
\setcounter{secnumdepth}{3}

\section{Generic effective potentials for microwave shielded molecules \label%
{sm:Veff}}

In the section, we derive an effective potential for microwave shielded
molecules. We focus on four lowest rotational states ($|J,M_{J}\rangle
=|0,0\rangle $, $|1,0\rangle $, and $|1,\pm 1\rangle $) of the molecule,
where the electric dipole moment of a molecule is $d\hat{\mathbf{d}}$ with $%
\hat{\mathbf{d}}$ being the unit vector along the internuclear axis of the
molecule. A position-independent elliptically polarized microwave field
propagating along the $z$ axis induces the transition between the rotational
ground state $|0,0\rangle $ and the excited state $|\xi _{+}\rangle \equiv
\cos \xi \left\vert 1,1\right\rangle +\sin \xi \left\vert 1,-1\right\rangle $%
, where $\xi $ is the elliptic angle. In the interaction picture, the
single-particle Hamiltonian reads%
\begin{equation}
\hat{h}_{\mathrm{in}}=\delta \sum_{M_{J}}|1,M_{J}\rangle \left\langle
1,M_{J}\right\vert +\frac{\Omega }{2}(|\xi _{+}\rangle \left\langle
0,0\right\vert +\mathrm{H.c.}),
\end{equation}%
where $\delta =\omega _{1}-\omega _{0}$ is the detuning between the energy
level spacing $\omega _{1}$ and the microwave frequency $\omega _{0}$, and $%
\Omega $ is the Rabi frequency. The eigenstates of $\hat{h}_{\mathrm{in}}$
are $|0\rangle \equiv |1,0\rangle $, $|\xi _{-}\rangle \equiv \cos \xi
\left\vert 1,-1\right\rangle -\sin \xi \left\vert 1,1\right\rangle $, $%
|+\rangle \equiv u|0,0\rangle +v|\xi _{+}\rangle $, and $|-\rangle \equiv
u|\xi _{+}\rangle -v|0,0\rangle $, where $u=\sqrt{(1-\delta /\Omega _{%
\mathrm{eff}})/2}$, $v=\sqrt{(1+\delta /\Omega _{\mathrm{eff}})/2}$, and $%
\Omega _{\mathrm{eff}}=\sqrt{\delta ^{2}+\Omega ^{2}}$ is the effective Rabi
frequency. The corresponding eigenenergies are $E_{0}=E_{\xi _{-}}=\delta $
and $E_{\pm }=(\delta \pm \Omega _{\mathrm{eff}})/2$.

In the center of mass frame, the two-body Hamiltonian is $H=-\nabla ^{2}/M+V(%
{\mathbf{r}})$, where ${\mathbf{r}}$ is the relative coordinate. The
interaction potential $V({\mathbf{r}})=V_{\mathrm{in}}-C_{\mathrm{vdW}%
}/r^{6} $ contains the internal-state dependent part $V_{\mathrm{in}%
}=\sum_{j=1,2}\hat{h}_{\mathrm{in},j}+V_{\mathrm{dd}}({\mathbf{r}})$ and the
universal van der Waals interaction described by $C_{\mathrm{vdW}}$, where%
\begin{eqnarray}
V_{\mathrm{dd}}({\mathbf{r}}) &=&\frac{d^{2}}{4\pi \epsilon _{0}r^{3}}\left[ 
\hat{\mathbf{d}}_{1}\cdot \hat{\mathbf{d}}_{2}-3(\hat{\mathbf{d}}_{1}\cdot 
\hat{\mathbf{r}})(\hat{\mathbf{d}}_{2}\cdot \hat{\mathbf{r}})\right]  \notag
\\
&=&-8\sqrt{\frac{2}{15}}\pi ^{3/2}\frac{d^{2}}{4\pi \epsilon _{0}r^{3}}%
\sum_{m=-2}^{2}Y_{2m}^{\ast }(\hat{\mathbf{r}})\Sigma _{2,m}
\end{eqnarray}%
is the dipolar interaction of molecules and $Y_{2m}(\hat{\mathbf{r}})$ are
spherical harmonics. The rank-2 spherical tensor $\Sigma _{2}$ with
components 
\begin{equation}
\Sigma _{2,0}=\frac{1}{\sqrt{6}}(\hat{d}_{1}^{+}\hat{d}_{2}^{-}+\hat{d}%
_{1}^{-}\hat{d}_{2}^{+}+2\hat{d}_{1}^{0}\hat{d}_{2}^{0}),\;\Sigma _{2,\pm 1}=%
\frac{1}{\sqrt{2}}(\hat{d}_{1}^{\pm }\hat{d}_{2}^{0}+\hat{d}_{1}^{0}\hat{d}%
_{2}^{\pm }),\;\mbox{and }\Sigma _{2,\pm 2}=\hat{d}_{1}^{\pm }\hat{d}%
_{2}^{\pm }
\end{equation}%
is determined by $\hat{d}_{j}^{\pm }=Y_{1,\pm 1}(\mathbf{\hat{d}}_{j})$ and $%
\hat{d}_{j}^{0}=Y_{1,0}(\mathbf{\hat{d}}_{j})$ with $\mathbf{\hat{d}}%
_{j=1,2} $ being the unit vector along the directions of the $j$th dipole
moment. In the basis $|J,M_{J}\rangle $,%
\begin{equation}
\hat{d}^{0}=\hat{d}_{p}^{0}e^{-i\omega _{0}t}+\hat{d}_{p}^{0\dagger
}e^{i\omega _{0}t},\;\hat{d}^{+}=\hat{d}_{p}^{+}e^{-i\omega _{0}t}+\hat{d}%
_{m}^{+}e^{i\omega _{0}t},\mbox{ and }\hat{d}^{-}=-(\hat{d}^{+})^{\dag },
\end{equation}%
where $\hat{d}_{p}^{0}=\left\vert 0,0\right\rangle \left\langle
1,0\right\vert /\sqrt{4\pi }$ and $\hat{d}_{p}^{+}=-\left\vert
0,0\right\rangle \left\langle 1,-1\right\vert /\sqrt{4\pi }$ are the
positive frequency parts, and $\hat{d}_{m}^{+}=\left\vert 1,1\right\rangle
\left\langle 0,0\right\vert /\sqrt{4\pi }$ is the negative frequency part.

For convenient, we study the scattering in the dressed-state basis $%
\{|+\rangle ,|0\rangle ,|\xi _{-}\rangle ,|-\rangle \}$, i.e., the
eigenbasis of $\hat{h}_{\mathrm{in},j}$. We focus on the microwave shielding
effect of two molecules in the highest dressed state $|1\rangle =|+,+\rangle 
$. It turns out that $V({\mathbf{r}})$ only couples $|1\rangle $ to other
six states $|2\rangle =|+,0\rangle _{s}$, $|3\rangle =|+,\xi _{-}\rangle
_{s} $, $|4\rangle =|+,-\rangle _{s}$, $|5\rangle =|-,0\rangle _{s}$, $%
|6\rangle =|-,\xi _{-}\rangle _{s}$, and $|7\rangle =|-,-\rangle $ in the
symmetric subspace, where $|i,j\rangle _{s}=(|i,j\rangle +|j,i\rangle )/%
\sqrt{2}$. Projecting $V({\mathbf{r}})$ in the subspace $\mathcal{S}%
_{7}=\{|\nu \rangle \}_{\nu =1}^{7}$, we obtain the potential%
\begin{equation}
V_{\nu \nu ^{\prime }}({\mathbf{r}})=(\mathcal{E}_{\nu }-\frac{C_{\mathrm{vdW%
}}}{r^{6}})\delta _{\nu \nu ^{\prime }}-8\sqrt{\frac{2}{15}}\pi ^{3/2}\frac{%
d^{2}}{4\pi \epsilon _{0}r^{3}}\sum_{m=-2}^{2}Y_{2m}^{\ast }(\hat{\mathbf{r}}%
)(\Sigma _{2,m})_{\nu \nu ^{\prime }},  \label{V7}
\end{equation}%
where the asymptotic energy%
\begin{equation}
\mathcal{E}=%
\begin{pmatrix}
0 & 0 & 0 & 0 & 0 & 0 & 0 \\ 
0 & \frac{1}{2}(\delta -\Omega _{\mathrm{eff}}) & 0 & 0 & 0 & 0 & 0 \\ 
0 & 0 & \frac{1}{2}(\delta -\Omega _{\mathrm{eff}}) & 0 & 0 & 0 & 0 \\ 
0 & 0 & 0 & -\Omega _{\mathrm{eff}} & 0 & 0 & 0 \\ 
0 & 0 & 0 & 0 & \frac{1}{2}(\delta -3\Omega _{\mathrm{eff}}) & 0 & 0 \\ 
0 & 0 & 0 & 0 & 0 & \frac{1}{2}(\delta -3\Omega _{\mathrm{eff}}) & 0 \\ 
0 & 0 & 0 & 0 & 0 & 0 & -2\Omega _{\mathrm{eff}}%
\end{pmatrix}%
\end{equation}%
is defined with respect to the highest channel. The spherical tensor $\Sigma
_{2}$ in the space $\mathcal{S}_{7}$ becomes%
\begin{align}
\Sigma _{2,0}& =\frac{1}{4\pi \sqrt{6}}%
\begin{pmatrix}
-2u^{2}v^{2} & 0 & 0 & -\sqrt{2}uvw & 0 & 0 & 2u^{2}v^{2} \\ 
0 & 2u^{2} & 0 & 0 & -2uv & 0 & 0 \\ 
0 & 0 & -u^{2} & 0 & 0 & uv & 0 \\ 
-\sqrt{2}uvw & 0 & 0 & -w^{2} & 0 & 0 & \sqrt{2}uvw \\ 
0 & -2uv & 0 & 0 & 2v^{2} & 0 & 0 \\ 
0 & 0 & uv & 0 & 0 & -v^{2} & 0 \\ 
2u^{2}v^{2} & 0 & 0 & \sqrt{2}uvw & 0 & 0 & -2u^{2}v^{2}%
\end{pmatrix}%
,  \notag \\
\Sigma _{2,1}& =\frac{1}{4\pi \sqrt{2}}%
\begin{pmatrix}
0 & \sqrt{2}u^{2}v\cos \xi & 0 & 0 & -\sqrt{2}uv^{2}\cos \xi & 0 & 0 \\ 
-\sqrt{2}u^{2}v\sin \xi & 0 & -u^{2}\cos \xi & -uw\sin \xi & 0 & uv\cos \xi
& \sqrt{2}u^{2}v\sin \xi \\ 
0 & -u^{2}\sin \xi & 0 & 0 & uv\sin \xi & 0 & 0 \\ 
0 & uw\cos \xi & 0 & 0 & -vw\cos \xi & 0 & 0 \\ 
\sqrt{2}uv^{2}\sin \xi & 0 & uv\cos \xi & vw\sin \xi & 0 & -v^{2}\cos \xi & -%
\sqrt{2}uv^{2}\sin \xi \\ 
0 & uv\sin \xi & 0 & 0 & -v^{2}\sin \xi & 0 & 0 \\ 
0 & -\sqrt{2}u^{2}v\cos \xi & 0 & 0 & \sqrt{2}uv^{2}\cos \xi & 0 & 0%
\end{pmatrix}%
,  \notag \\
\Sigma _{2,2}& =-\frac{1}{4\pi }%
\begin{pmatrix}
u^{2}v^{2}\sin 2\xi & 0 & \sqrt{2}u^{2}v\cos ^{2}\xi & \frac{\sqrt{2}}{2}%
uvw\sin 2\xi & 0 & -\sqrt{2}uv^{2}\cos ^{2}\xi & -u^{2}v^{2}\sin 2\xi \\ 
0 & 0 & 0 & 0 & 0 & 0 & 0 \\ 
-\sqrt{2}u^{2}v\sin ^{2}\xi & 0 & -\frac{1}{2}u^{2}\sin 2\xi & -uw\sin
^{2}\xi & 0 & \frac{1}{2}uv\sin 2\xi & \sqrt{2}u^{2}v\sin ^{2}\xi \\ 
\frac{\sqrt{2}}{2}uvw\sin 2\xi & 0 & uw\cos ^{2}\xi & \frac{1}{2}w^{2}\sin
2\xi & 0 & -vw\cos ^{2}\xi & -\frac{\sqrt{2}}{2}uvw\sin 2\xi \\ 
0 & 0 & 0 & 0 & 0 & 0 & 0 \\ 
\sqrt{2}uv^{2}\sin ^{2}\xi & 0 & \frac{1}{2}uv\sin 2\xi & vw\sin ^{2}\xi & 0
& -\frac{1}{2}v^{2}\sin 2\xi & -\sqrt{2}uv^{2}\sin ^{2}\xi \\ 
-u^{2}v^{2}\sin 2\xi & 0 & -\sqrt{2}u^{2}v\cos ^{2}\xi & -\frac{\sqrt{2}}{2}%
uvw\sin 2\xi & 0 & \sqrt{2}uv^{2}\cos ^{2}\xi & u^{2}v^{2}\sin 2\xi%
\end{pmatrix}%
,
\end{align}%
where $w=u^{2}-v^{2}$.

The adiabatic potential of the highest channel can be obtained by
diagonalizing the $7\times 7$ matrix $V_{\nu \nu ^{\prime }}({\mathbf{r}})$
for all ${\mathbf{r}}$. The largest eigenvalue $V_{\mathrm{adia}}({\mathbf{r}%
})$ gives rise to the effective potential for two incident molecules in the
state $|1\rangle $. The corresponding position dependent eigen-vector $|1({%
\mathbf{r}})\rangle =\sum_{\nu }\alpha _{\nu }({\mathbf{r}})|\nu \rangle $
describes the mixing of the seven bare channels induced by the dipolar
interaction, where $\alpha _{\nu }({\mathbf{r\rightarrow \infty }}%
)=(1,0,0,0,0,0,0)^{T}$. For the low energy scattering, the second order
perturbation leads to the effective potential $V_{\mathrm{adia}}({\mathbf{r}}%
)\sim V_{\mathrm{eff}}(\mathbf{r})$:%
\begin{eqnarray}
V_{\mathrm{eff}}(\mathbf{r}) &=&\frac{C_{6}}{r^{6}}\sin ^{2}\theta \{1-%
\mathcal{F}_{\xi }^{2}(\varphi )+[1-\mathcal{F}_{\xi }(\varphi )]^{2}\cos
^{2}\theta \}  \notag \\
&&+\frac{C_{3}}{r^{3}}[3\cos ^{2}\theta -1+3\mathcal{F}_{\xi }(\varphi )\sin
^{2}\theta ]-\frac{C_{\mathrm{vdW}}}{r^{6}},
\end{eqnarray}%
where $\mathcal{F}_{\xi }(\varphi )=\sin 2\xi \cos 2\varphi $, $\theta $ and 
$\varphi $ are the polar and azimuthal angles of ${\mathbf{r}}$. The
strength $C_{3}=d^{2}/\left[ 48\pi \epsilon _{0}(1+\delta _{r}^{2})\right] $
of the long range DDI only depends on the relative detuning $\delta
_{r}=|\delta |/\Omega $, and $C_{6}>0$ is obtained by fitting with the
adiabatic potential $V_{\mathrm{adia}}$.

\begin{figure}[tbp]
\includegraphics[width=0.9\linewidth]{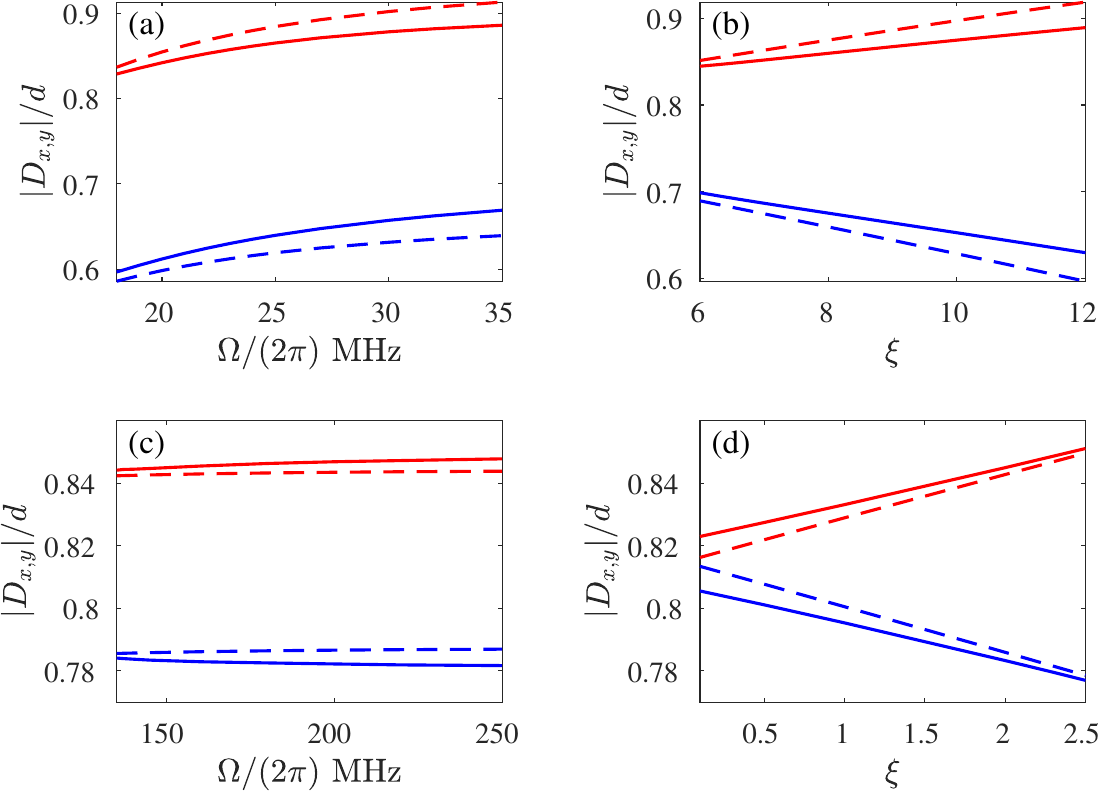} 
\caption{Dipole moments $\left\vert D_{x,y}\right\vert$ of the $p_y$- and $p_x$-tetramer states are shown in the first and second rows, respectively, where $\protect\delta/(2\pi)=-9.5\mathrm{MHz}$. The solid blue and red curves represent $\left\vert D_{x}\right\vert$ and $\left\vert D_{y}\right\vert$, while the dashed blue and red curves show
the dipole moments $\left\vert D_{\mathrm{tot},x}\right\vert$ and $\left\vert D_{\mathrm{tot},y}\right\vert$ of two widely separated molecules. (a) $\protect\xi=10^{\circ}$; (b) $\Omega/(2\pi)=28.5\mathrm{MHz}$; (c) $\protect\xi=2^{\circ}$; (d) $\Omega/(2\pi)=150\mathrm{MHz}$.}
\label{dipmom}
\end{figure}

Since the shielding core around $10^{2}\sim 10^{3}a_{0}$ generated by the $%
C_{6}$ potential is much larger than the range of the van der Waals
potential, two shielded\ molecules in the tetramer bound state with distance 
$10^{2}\sim 10^{3}a_{0}$ barely feel the van der Waals potential. Therefore,
we study the binding energy of the\ tetramer state by neglecting the van der
Waals $C_{\mathrm{vdW}}$-term. However, we note that the $C_{\mathrm{vdW}}$%
-term has the effect on the formation of the four body complex inside the
shielding core, thus, the life-time of the tetramer state is calculated by
taking into account the van der Waals potential. Using the B-spline
algorithm~\cite{Bachau_2001}, we numerically solve the Schr\"{o}dinger
equation (SE)%
\begin{equation}
\lbrack -\frac{\nabla ^{2}}{M}+V_{\mathrm{eff}}({\mathbf{r}})]\psi _{B}({%
\mathbf{r}})=\varepsilon _{B}\psi _{B}({\mathbf{r}})
\end{equation}%
to determine the tetramer wavefunction $\psi _{B}({\mathbf{r}})$ and the
binding energy $\left\vert \varepsilon _{B}\right\vert $.

In the seven bare dressed-state basis, the tetramer state is $\Psi _{B}({%
\mathbf{r}})=\sum_{\nu }\psi _{B}({\mathbf{r}})\alpha _{\nu }({\mathbf{r}}%
)|\nu \rangle $. The dipole moment of the tetramer is $\mathbf{D}=d\sum_{j=1,2}\int d{\mathbf{r}}\Psi _{B}^{\dagger }({\mathbf{r}})%
\mathbf{\hat{d}}_{j}\Psi _{B}({\mathbf{r}})$. In the explicit form,%
\begin{eqnarray}
\mathbf{D}^{z} &=&4d\sqrt{\frac{\pi }{3}}\sum_{j=1,2}\text{%
Re}\left\langle \hat{d}_{p,j}^{0}\right\rangle e^{-i\omega _{0}t},  \notag \\
\mathbf{D}^{x}&=&-2d\sqrt{\frac{2\pi }{3}}\sum_{j=1,2}%
\text{Re}(\left\langle \hat{d}_{p,j}^{+}\right\rangle +\left\langle \hat{d}%
_{m,j}^{+\dagger }\right\rangle )e^{-i\omega _{0}t}  \notag \\
&=&-2d\sqrt{\frac{2\pi }{3}}\sum_{j=1,2}[\text{Re}(\left\langle \hat{d}%
_{p,j}^{+}\right\rangle +\left\langle \hat{d}_{m,j}^{+}\right\rangle )\cos
\omega _{0}t+\text{Im}(\left\langle \hat{d}_{p,j}^{+}\right\rangle
-\left\langle \hat{d}_{m,j}^{+}\right\rangle )\sin \omega _{0}t] \\
\mathbf{D}^{y} &=&2d\sqrt{\frac{2\pi }{3}}\sum_{j=1,2}\text{%
Re}i(\left\langle \hat{d}_{p,j}^{+}\right\rangle -\left\langle \hat{d}%
_{m,j}^{+\dagger }\right\rangle )e^{-i\omega _{0}t}  \notag \\
&=&-2d\sqrt{\frac{2\pi }{3}}\sum_{j=1,2}[\text{Im}(\left\langle \hat{d}%
_{p,j}^{+}\right\rangle +\left\langle \hat{d}_{m,j}^{+}\right\rangle )\cos
\omega _{0}t-\text{Re}(\left\langle \hat{d}_{p,j}^{+}\right\rangle
-\left\langle \hat{d}_{m,j}^{+}\right\rangle )\sin \omega _{0}t]
\end{eqnarray}%
where the average value is defined as%
\begin{equation}
\left\langle O\right\rangle =\int d{\mathbf{r}}\left\vert \psi _{B}({\mathbf{%
r}})\right\vert ^{2}\sum_{\nu \nu ^{\prime }}\alpha _{\nu ^{\prime }}^{\ast
}({\mathbf{r}})\left\langle \nu ^{\prime }\right\vert O|\nu \rangle \alpha
_{\nu }({\mathbf{r}}).
\end{equation}%
Since $\mathcal{S}_{7}$ is the symmetric subspace, we only have to calculate
the average values for one molecule. For instance, the matrices for the
molecule $j=1$ are%
\begin{eqnarray}
\hat{d}_{p,1}^{0} &=&\frac{1}{2\sqrt{2\pi }}%
\begin{pmatrix}
0 & u & 0 & 0 & 0 & 0 & 0 \\ 
0 & 0 & 0 & 0 & 0 & 0 & 0 \\ 
0 & 0 & 0 & 0 & 0 & 0 & 0 \\ 
0 & -\frac{v}{\sqrt{2}} & 0 & 0 & \frac{u}{\sqrt{2}} & 0 & 0 \\ 
0 & 0 & 0 & 0 & 0 & 0 & 0 \\ 
0 & 0 & 0 & 0 & 0 & 0 & 0 \\ 
0 & 0 & 0 & 0 & -v & 0 & 0%
\end{pmatrix}%
,  \notag \\
\hat{d}_{p,1}^{+} &=&%
\begin{pmatrix}
-\frac{uv\sin \xi }{2\sqrt{\pi }} & 0 & -\frac{u\cos \xi }{2\sqrt{2\pi }} & -%
\frac{u^{2}\sin \xi }{2\sqrt{2\pi }} & 0 & 0 & 0 \\ 
0 & -\frac{uv\sin \xi }{4\sqrt{\pi }} & 0 & 0 & -\frac{u^{2}\sin \xi }{4%
\sqrt{\pi }} & 0 & 0 \\ 
0 & 0 & -\frac{uv\sin \xi }{4\sqrt{\pi }} & 0 & 0 & -\frac{u^{2}\sin \xi }{4%
\sqrt{\pi }} & 0 \\ 
\frac{v^{2}\sin \xi }{2\sqrt{2\pi }} & 0 & \frac{v\cos \xi }{4\sqrt{\pi }} & 
0 & 0 & -\frac{u\cos \xi }{4\sqrt{\pi }} & -\frac{u^{2}\sin \xi }{2\sqrt{%
2\pi }} \\ 
0 & \frac{v^{2}\sin \xi }{4\sqrt{\pi }} & 0 & 0 & \frac{uv\sin \xi }{4\sqrt{%
\pi }} & 0 & 0 \\ 
0 & 0 & \frac{v^{2}\sin \xi }{4\sqrt{\pi }} & 0 & 0 & \frac{uv\sin \xi }{4%
\sqrt{\pi }} & 0 \\ 
0 & 0 & 0 & \frac{v^{2}\sin \xi }{2\sqrt{2\pi }} & 0 & \frac{v\cos \xi }{2%
\sqrt{2\pi }} & \frac{uv\sin \xi }{2\sqrt{\pi }}%
\end{pmatrix}%
,  \notag \\
\hat{d}_{m,1}^{+} &=&%
\begin{pmatrix}
\frac{uv\cos \xi }{2\sqrt{\pi }} & 0 & 0 & -\frac{v^{2}\cos \xi }{2\sqrt{%
2\pi }} & 0 & 0 & 0 \\ 
0 & \frac{uv\cos \xi }{4\sqrt{\pi }} & 0 & 0 & -\frac{v^{2}\cos \xi }{4\sqrt{%
\pi }} & 0 & 0 \\ 
-\frac{u\sin \xi }{2\sqrt{2\pi }} & 0 & \frac{uv\cos \xi }{4\sqrt{\pi }} & 
\frac{v\sin \xi }{4\sqrt{\pi }} & 0 & -\frac{v^{2}\cos \xi }{4\sqrt{\pi }} & 
0 \\ 
\frac{u^{2}\cos \xi }{2\sqrt{2\pi }} & 0 & 0 & 0 & 0 & 0 & -\frac{v^{2}\cos
\xi }{2\sqrt{2\pi }} \\ 
0 & \frac{u^{2}\cos \xi }{4\sqrt{\pi }} & 0 & 0 & -\frac{uv\cos \xi }{4\sqrt{%
\pi }} & 0 & 0 \\ 
0 & 0 & \frac{u^{2}\cos \xi }{4\sqrt{\pi }} & -\frac{u\sin \xi }{4\sqrt{\pi }%
} & 0 & -\frac{uv\cos \xi }{4\sqrt{\pi }} & \frac{v\sin \xi }{2\sqrt{2\pi }}
\\ 
0 & 0 & 0 & \frac{u^{2}\cos \xi }{2\sqrt{2\pi }} & 0 & 0 & -\frac{uv\cos \xi 
}{2\sqrt{\pi }}%
\end{pmatrix}%
.
\end{eqnarray}

It turns out that $\mathbf{D}^{z}=0$, $\mathbf{D}^{x}=D_{x}\cos \omega _{0}t$, and $\mathbf{D}^{y}=D_{y}\sin \omega _{0}t$, where%
\begin{eqnarray}
D_{x} &=&-2d\sqrt{\frac{2\pi }{3}}\sum_{j=1,2}(\left\langle \hat{d}%
_{p,j}^{+}\right\rangle +\left\langle \hat{d}_{m,j}^{+}\right\rangle ), 
\notag \\
D_{y} &=&2d\sqrt{\frac{2\pi }{3}}\sum_{j=1,2}(\left\langle \hat{d}%
_{p,j}^{+}\right\rangle -\left\langle \hat{d}_{m,j}^{+}\right\rangle ).
\end{eqnarray}%
In Fig.~\ref{dipmom}, we show $\left\vert D_{x}\right\vert $ and $\left\vert
D_{y}\right\vert $ as the function of $\Omega $ and $\xi $, where the sum $\mathbf{D}_{\rm{tot}}\equiv d\sum_{j}\left\langle +,+\right\vert \hat{%
\mathbf{d}}_{j}\left\vert +,+\right\rangle $ of dipole
moments for two widely separated molecules is also plotted as a reference.

\section{Coupled-channel scattering calculations \label{sm:CC}}

In this section, we perform the multi-channel scattering calculations to
obtain the accurate binding energy and the life-time of the tetramer state.
The multi-channel SE reads%
\begin{equation}
\sum_{\nu ^{\prime }=1}^{7}\left( -\frac{\nabla ^{2}}{M}\delta _{\nu \nu
^{\prime }}+V_{\nu \nu ^{\prime }}\right) \psi _{\nu ^{\prime }}({\mathbf{r}}%
)=E\psi _{\nu }({\mathbf{r}}),  \label{SE7}
\end{equation}%
where $E=k^{2}/M+\mathcal{E}_{\nu _{0}}$ is the incident energy of the $\nu
_{0}$th channel. We expand $\psi _{\nu }({\mathbf{r}})=\sum_{lm}r^{-1}\phi
_{\nu lm}Y_{lm}(\hat{\mathbf{r}})$ in the angular momentum basis, where $l$
is even (odd) for bosons (fermions). The Schr\"{o}dinger equation for $\phi
_{\nu lm}$ follows from Eq. (\ref{SE7}) as%
\begin{equation}
\sum_{\nu ^{\prime }l^{\prime }m^{\prime }}\left[ \left( -\frac{\partial ^{2}%
}{\partial r^{2}}+\frac{l(l+1)}{r^{2}}-\frac{MC_{\mathrm{vdW}}}{r^{6}}%
\right) \delta _{\nu \nu ^{\prime }}\delta _{ll^{\prime }}\delta
_{mm^{\prime }}-\frac{Md^{2}}{4\pi \epsilon _{0}r^{3}}\sum_{s=-2}^{2}(\Xi
_{s})_{\nu lm,\nu ^{\prime }l^{\prime }m^{\prime }}\right] \phi _{\nu
^{\prime }l^{\prime }m^{\prime }}(r)=k_{\nu }^{2}\phi _{\nu lm}(r),
\label{SECC}
\end{equation}%
where $k_{\nu }=\sqrt{k^{2}+M(\mathcal{E}_{\nu _{0}}-\mathcal{E}_{\nu })}$
and%
\begin{equation}
(\Xi _{s})_{\nu lm,\nu ^{\prime }l^{\prime }m^{\prime }}=4\pi \sqrt{\frac{%
2(2l^{\prime }+1)}{3(2l+1)}}C_{l^{\prime }020}^{l0}C_{l^{\prime }m^{\prime
}2s}^{lm}(\Sigma _{2,s}^{\dagger })_{\nu \nu ^{\prime }}
\end{equation}
$C_{l^{\prime }m^{\prime }l^{\prime \prime }m^{\prime \prime }}^{lm}$ being
the short-hand notation for the Clebsch-Gordan coefficients. In the compact
form, Eq. (\ref{SECC}) reads%
\begin{equation}
\left[ \frac{\partial ^{2}}{\partial r^{2}}+\mathcal{V}(r)\right] \phi (r)=0,
\label{cSE}
\end{equation}%
where the potential%
\begin{equation}
\mathcal{V}(r)=\left[ k_{\nu }^{2}-\frac{l(l+1)}{r^{2}}+\frac{MC_{\mathrm{vdW%
}}}{r^{6}}\right] \delta _{\nu \nu ^{\prime }}\delta _{ll^{\prime }}\delta
_{mm^{\prime }}+\frac{Md^{2}}{4\pi \epsilon _{0}r^{3}}\sum_{s=-2}^{2}(\Xi
_{s})_{\nu lm,\nu ^{\prime }l^{\prime }m^{\prime }}.
\end{equation}%
The coupled channel SE (\ref{cSE}) can be solved using the log-derivative
method \cite{log-Johnson} with high precision. To describe the formation of
the four body complex inside the shielding core, we employ the capture
boundary condition~\cite{Clary1987,Rackham2003} at $r=r_{0}$. We note that
since the wavefunction has a tiny distribution inside the shielding core the
result is not affected by the choice of $r_{0}<10^{2}a_{0}$. In the
numerical calculation, the result is stable for different choices $r_{0}=32a_{0}$, $48.5a_{0}$, and $64a_{0}$.

\begin{figure}[tbp]
\includegraphics[width=0.9\linewidth]{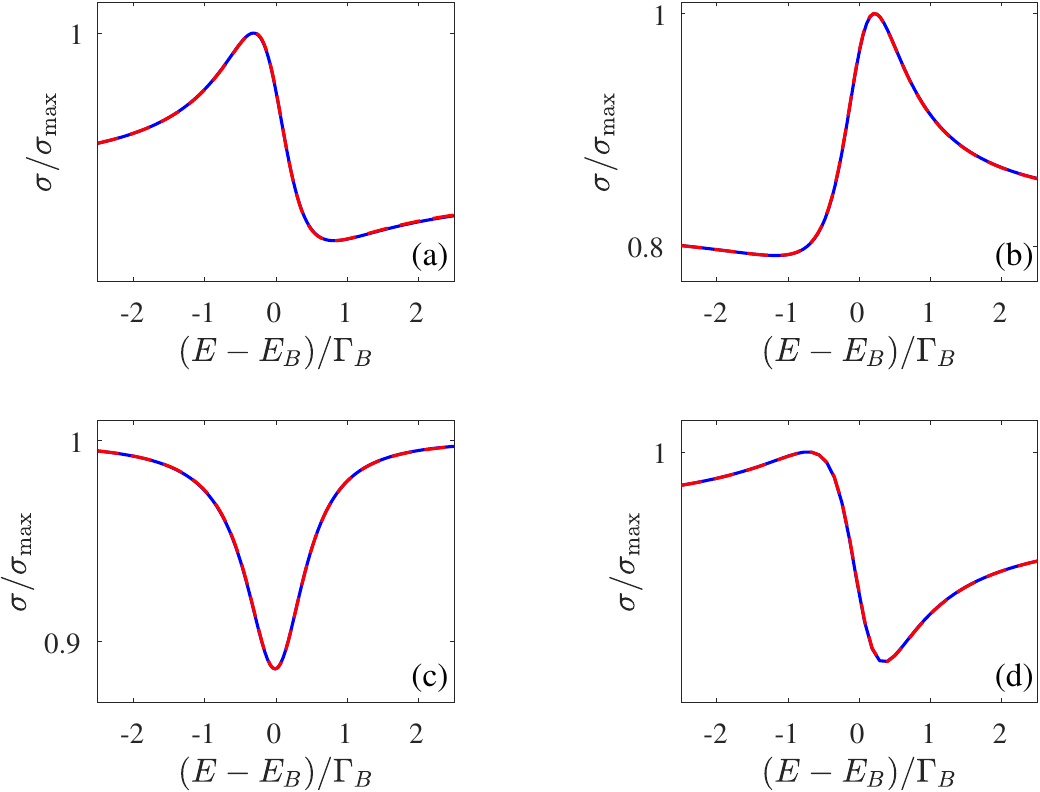}  
\caption{Comparisons of lineshapes obtained from coupled channel
calculations and Eq. (2) in the main text, where $\protect\delta/(2\pi)=-9.5\mathrm{MHz}$: (a) $(\Omega/(2\pi),\protect\xi)=(19.52\ \mathrm{MHz},10^{\circ})$ and (b) $(\Omega/(2\pi),\protect\xi)=(61.58\ \mathrm{MHz},2^{\circ})$ display the lineshapes around the
resonance with the $p_y$-tetramer state $\left\vert \protect\psi %
_{B,+}\right\rangle$; (c) $(\Omega/(2\pi),\protect\xi)=(148.94\ \textrm{MHz},2^{\circ})$ and (d) $(\Omega/(2\pi),\protect\xi)=(243.59\ \textrm{MHz},4^{\circ})$ display the lineshapes around the
resonance with the $p_x$-tetramer state $\left\vert \protect\psi %
_{B,-}\right\rangle$.}
\label{lineshapecomp}
\end{figure}

By matching the numerical solution with the asymptotic wavefunction, we can
obtain the scattering $K$-matrix, which results in the scattering amplitude%
\begin{equation}
f_{\nu lm}^{\nu ^{\prime }l^{\prime }m^{\prime }}=i\frac{1}{\sqrt{k_{\nu
^{\prime }}}}\left( \frac{1}{K+i}K\right) _{\nu ^{\prime }l^{\prime
}m^{\prime },\nu lm}\frac{1}{\sqrt{k_{\nu }}}
\end{equation}%
from the channel $(\nu lm)$ to the channel $(\nu ^{\prime }l^{\prime
}m^{\prime })$, and the partial wave scattering cross section $\sigma _{\nu
lm}^{\nu ^{\prime }l^{\prime }m^{\prime }}=4\pi \left\vert f_{\nu lm}^{\nu
^{\prime }l^{\prime }m^{\prime }}\right\vert ^{2}$. We can also obtain the
average elastic and inelastic scattering cross sections $\sigma _{\nu }^{%
\mathrm{el}}=\sum_{lml^{\prime }m^{\prime }}\sigma _{\nu lm}^{\nu ^{\prime
}l^{\prime }m^{\prime }}$ and $\sigma _{\nu }^{\mathrm{inel}}=\sum_{\nu
^{\prime },lml^{\prime }m^{\prime }}k_{\nu ^{\prime }}\sigma _{\nu lm}^{\nu
^{\prime }l^{\prime }m^{\prime }}/k_{\nu }$.

In Fig.~\ref{lineshapecomp}, we show the scattering cross sections $\sigma
_{210}^{210}$, which display perfect agreement with the analytical
expression (2) in the main text for the Feshbach resonances. We show the
binding energy $|E_{B}|$ and the decay rate $\Gamma_{B}$ of the tetramer
state extracted from the lineshape in the main text. Here, in Fig.~\ref%
{EBcomp} we present the comparison between $|E_{B}|$ and the binding energy $%
\varepsilon _{B}$ obtained from the effective potential.

We note that the lifetime of the $p_x$-tetramer state for LiBb molecules can be extended to 10.4s for the system parameters $\xi=0.5$, $\delta=2\pi \times 1$MHz, and $\Omega=2\pi \times 98.3$MHz.

\begin{figure}[tbp]
\includegraphics[width=1.0\linewidth]{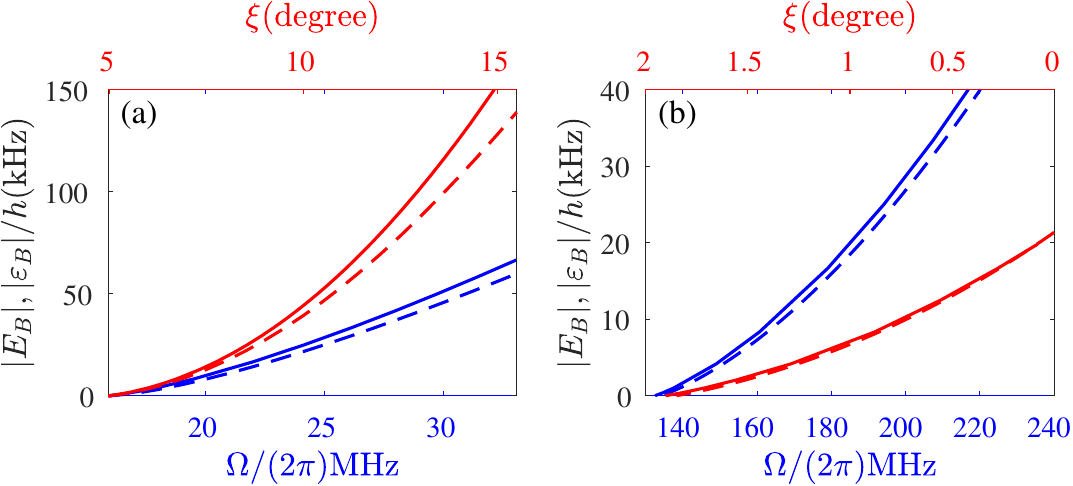} 
\caption{Comparisons of binding energies obtained from fitting lineshapes to
Eq. (2) in the main text (solid curves) and the effective potential (dashed
curves), where $\protect\delta/(2\pi)=-9.5\mathrm{MHz}$. (a) The blue and red curves display
binding energies of the $p_y$-tetramer state along the horizontal and vertical cuts in Fig.
1c of the main text; (b) The blue and red curves display
binding energies of the $p_x$-tetramer state along the horizontal and vertical cuts in Fig.
1d of the main text. }
\label{EBcomp}
\end{figure}

\section{Tetramer Associations and Dissociations \label{Associations}}

The tetramer association can be simulated by evolving the time-dependent SE
corresponding to the effective Hamiltonian $H_{\mathrm{eff}}(t)$, where $%
C_{3}$, $C_{6}$, and $\mathcal{F}_{\xi }$ are time-dependent. The conversion
efficiency ~\cite{Burnett2007}%
\begin{equation}
\mathcal{E}_{A}=2n{\lambda _{T}^{3}}\int d\mathbf{k}e^{-\beta 
\frac{k^{2}}{M}}p_{\mathrm{a}}(\mathbf{k})
\end{equation}%
to the tetramer can be evaluated using the density $n$ of molecular gases
and the thermal de Broglie wavelength $\lambda _{T}=(4\pi \hbar^2 /MT)^{1/2}$ at
temperature $T$, where the transition probability $p_{\mathrm{a}}(\mathbf{k}%
)=\left\vert \left\langle \psi _{B}\right\vert {U(\tau _{\mathrm{a}})}%
\left\vert \psi _{\mathbf{k}}\right\rangle \right\vert ^{2}$ in the
association time $\tau _{\mathrm{a}}$ is determined by the unitary evolution
operator {$U(\tau _{\mathrm{a}})=$}$\mathcal{T}\exp [-i\int_{0}^{\tau _{%
\mathrm{a}}}dtH_{\mathrm{eff}}(t)]$. Here, the initial state $\left\vert
\psi _{\mathbf{k}}\right\rangle $ is the scattering eigenstate of $H_{%
\mathrm{eff}}(0)$, and the final state $\left\vert \psi _{B}\right\rangle $
is the tetramer state of $H_{\mathrm{eff}}(\tau _{a})$.

\begin{figure}[tbp]
\includegraphics[width=1.0\linewidth]{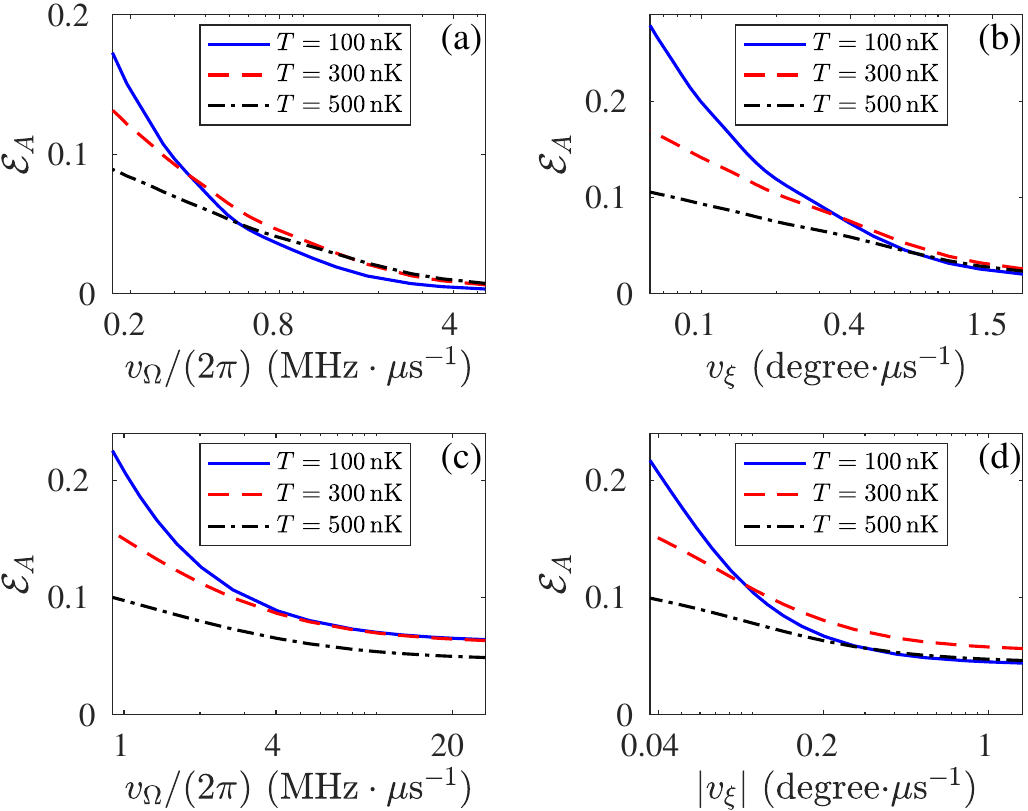}
\caption{Association efficiency as a function of ramp speeds $v_{\Omega}=\partial_t \Omega$ and $v_{\xi}=\partial_t \xi$, respectively. (a) For $\xi=10^{\circ}$, the ramp of $\Omega/(2\pi)$ from $5\mathrm{MHz}$ to $35\mathrm{MHz}$ with speed $v_{\Omega}$; (b) For $\Omega/(2\pi)=28.5\mathrm{MHz}$, the ramp of $\xi$ from $1^{\circ}$ to $12^{\circ}$ with speed $v_{\xi}$; (c) For $\xi=2^{\circ}$, the ramp of $\Omega/(2\pi)$ from $80\mathrm{MHz}$ to $170\mathrm{MHz}$ with speed $v_{\Omega}$; (d) For $\Omega/(2\pi)=150\mathrm{MHz}$, the ramp of $\xi$ from $5^{\circ}$ to $0.5^{\circ}$ with speed $v_{\xi}$.}
\label{asso}
\end{figure}

In the left and right panels of Fig.~\ref{asso}, we show the conversion
efficiency $\mathcal{E}_{A}$ as a function of ramp speeds $v_{\Omega}=\partial_t \Omega$ and
$v_{\xi}=\partial_t \xi$, respectively. In the first and second rows, $\mathcal{E}_{A}$ for the $p_y$- and $p_x$-tetramer states are plotted. Adiabaticity is
crucial for high conversion efficiency, and slower ramps lead to higher
adiabaticity. Additionally, for slow ramp speeds, the conversion efficiency to tetramer states is higher at lower temperatures $T$. It should be noted that this result, obtained from two-body
scattering calculations, is not valid in the saturation regime of $\mathcal{E%
}_{A}$~\cite{Williams2006}, where many-body effects must be considered using the non-Markovian Boltzmann equation.

A reverse fast ramp leads to the dissociation of tetramers initially at the
temperature $T$. After the dissociation process, the wavefunction $\psi _{%
\mathrm{d}}(\mathbf{k})=\left\langle \mathbf{k}\right\vert {U(\tau )}%
\left\vert \psi _{B}^{(x,y)}\right\rangle $ in the center of mass frame is
obtained using the B-spline method, where $\mathbf{k}$ is the relative
momentum. The momentum distribution of the molecule dissociated from the
tetramer is
\begin{eqnarray}
n_{T}(\mathbf{k}) &=&\frac{1}{(4\pi MT)^{3/2}}\int d\mathbf{p}e^{-\beta \frac{(%
\mathbf{k}+\mathbf{p})^{2}}{4M}}\left\vert \psi _{\mathrm{d}}(\frac{\mathbf{k%
}-\mathbf{p}}{2})\right\vert ^{2}  \notag \\
&=&\frac{8}{(4\pi MT)^{3/2}}\int d\mathbf{p}e^{-\beta \frac{(\mathbf{k}-%
\mathbf{p})^{2}}{M}}\left\vert \psi _{\mathrm{d}}(\mathbf{p})\right\vert ^{2}
\notag \\
&=&\int d\mathbf{r}e^{-i\mathbf{k}\cdot \mathbf{r}}e^{-\frac{M}{4\beta }%
\mathbf{r}^{2}}\int \frac{d\mathbf{p}}{(2\pi )^{3}}e^{i\mathbf{p}\cdot
\mathbf{r}}\left\vert \psi _{\mathrm{d}}(\mathbf{p})\right\vert ^{2}  \notag
\\
&=&\frac{2}{\pi }\sum_{lm}Y_{lm}(\hat{k})\int r^{2}dre^{-\frac{M}{4\beta }%
r^{2}}j_{l}(kr)\int p^{2}dpj_{l}(pr)\int d\Omega _{p}Y_{lm}^{\ast }(\hat{p}%
)\left\vert \psi _{\mathrm{d}}(\mathbf{p})\right\vert ^{2}.
\end{eqnarray}%

The energy distributions reads%
\begin{eqnarray}
f_{T}(E) &=&\int d\mathbf{k}\delta (E-\frac{k^{2}}{2M})n_{T}(\mathbf{k})  \notag \\
&=&\int d\mathbf{K}d\mathbf{k}\delta (E-\frac{1}{2M}(\frac{\mathbf{K}}{2}+%
\mathbf{k})^{2})\frac{1}{(4\pi MT)^{3/2}}e^{-\beta \frac{K^{2}}{4M}%
}\left\vert \psi _{\mathrm{d}}(\mathbf{k})\right\vert ^{2}  \notag \\
&=&2\pi \int d\mathbf{k}\left\vert \psi _{\mathrm{d}}(\mathbf{k})\right\vert
^{2}\int K^{2}dK\int d\cos \theta _{K}\delta (E-\frac{1}{2M}(\frac{K^{2}}{4}%
+k^{2})-\frac{Kk}{2M}\cos \theta _{K})\frac{e^{-\beta \frac{K^{2}}{4M}}}{%
(4\pi MT)^{3/2}}  \notag \\
&=&\int d\mathbf{k}\left\vert \psi _{\mathrm{d}}(\mathbf{k})\right\vert ^{2}%
\frac{M}{k}4\pi \int KdK\frac{e^{-\beta \frac{K^{2}}{4M}}}{(4\pi MT)^{3/2}}%
\theta (E-\frac{1}{2M}(\frac{K}{2}-k)^{2})\theta (\frac{1}{2M}(\frac{K}{2}%
+k)^{2}-E)  \notag \\
&=&\int d\mathbf{k}\left\vert \psi _{\mathrm{d}}(\mathbf{k})\right\vert ^{2}%
\frac{M}{k(\pi MT)^{1/2}}[e^{-\frac{\beta }{M}(k-\sqrt{2ME})^{2}}-e^{-\frac{%
\beta }{M}(k+\sqrt{2ME})^{2}}].
\end{eqnarray}

\begin{figure}[tbp]
\includegraphics[width=1.0\linewidth]{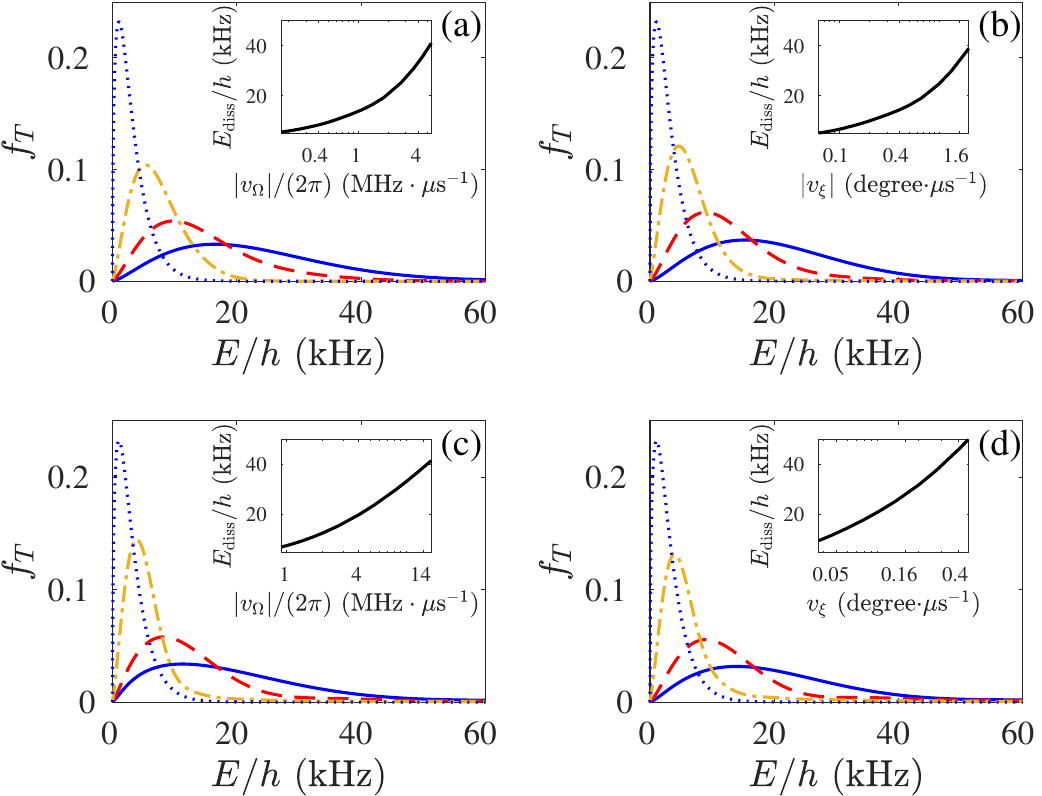}
\caption{Ramp dissociations for the the $p_y$- and $p_x$-tetramers (in the
first and second rows) with different ramp speeds, alongside the thermal
distribution at $T=100$nK for reference (dashed blue curves). (a) For $%
\protect\xi=10^{\circ}$, the ramp of $\Omega/(2\protect\pi)$ from $35\mathrm{%
MHz}$ to $5\mathrm{MHz}$ with speed $v_{\Omega}/(2\protect\pi)=0.45\mathrm{%
MHz}/\protect\mu s$ (the yellow dash-dotted curve),$1.36\mathrm{MHz}/\protect\mu s$ (the
red dashed curve), and $2.73\mathrm{MHz}/\protect\mu s$ (the blue solid curve); (b) For $%
\Omega/(2\protect\pi)=28.5\mathrm{MHz}$, the ramp of $\protect\xi$ from $%
12^{\circ}$ to $1^{\circ}$ with speed $v_{\protect\xi}=0.17^{\circ}/\protect%
\mu s$ (the yellow dash-dotted curve), $0.5^{\circ}/\protect\mu s$ (the red dashed curve), and $%
1^{\circ}/\protect\mu s$ (the blue solid curve); (c) For $\protect\xi=2^{\circ}$,
the ramp of $\Omega/(2\protect\pi)$ from $170\mathrm{MHz}$ to $80\mathrm{MHz}
$ with speed $v_{\Omega}/(2\protect\pi)=1.36\mathrm{MHz}/\protect\mu s$ (the yellow dash-dotted curve),$4.09\mathrm{MHz}/\protect\mu s$ (the red dashed curve), and $8.18%
\mathrm{MHz}/\protect\mu s$ (the blue solid curve); (d) For $\Omega/(2\protect\pi%
)=150\mathrm{MHz}$, the ramp of $\protect\xi$ from $0.5^{\circ}$ to $%
5^{\circ}$ with speed $v_{\protect\xi}=0.07^{\circ}/\protect\mu s$ (the yellow dash-dotted curve), $0.2^{\circ}/\protect\mu s$ (the red dashed curve), and $0.41^{\circ}/%
\protect\mu s$ (the blue solid curve). The insets show $E_\mathrm{diss}$ as
functions of ramp speeds. The thermal distribution at $T=100\mathrm{nK}$ is
shown by the dashed blue curve for reference.}
\label{diss}
\end{figure}

In Fig.~\ref{diss}, for the $p_{y}$- and $p_{x}$-tetramers, we show the
energy distributions $f_{T}(E)$ in final states of ramp dissociations with
different ramp speeds, where the thermal distribution at $T=100$nK is
plotted for reference. In insets, the dissociation energy $E_{\mathrm{diss}}$
is displayed. The energy distributions via ramps of $\Omega $ and $\xi $
exhibit a similar structure.

\section{Effective Hamiltonian for modulational dissociations \label{MD}}

In this section, we derive the effective Hamiltonian to describe the
dissociation under small modulations of the elliptic angle. For two molecules in the highest adiabatic
channel, the system is described by $H_{\mathrm{eff}}(\xi )=-\nabla
^{2}/M+V_{\mathrm{eff}}(\mathbf{r})$. When the elliptic angle $\xi $ is
modulated, $C_{6}$ and $\mathcal{F}_{\xi }(\varphi )$ in $V_{\mathrm{eff}}(%
\mathbf{r})$, which depend on $\xi $, are both time-dependent. For a small
modulation ${\xi }=\xi _{0}+\delta \xi \sin \omega _{\mathrm{m}}t$ around
the background value $\xi _{0}$, the Taylor expansion leads to%
\begin{equation}
H_{\mathrm{eff}}(\xi )\sim H_{\mathrm{eff}}(\xi _{0})+\delta \xi \partial
_{\xi _{0}}V_{\mathrm{eff}}(\mathbf{r})\sin \omega _{\mathrm{m}}t,
\end{equation}%
where in the explicit form%
\begin{eqnarray}
\partial _{\xi _{0}}V_{\mathrm{eff}}(\mathbf{r}) &=&6\frac{C_{3}}{r^{3}}\cos
2\xi _{0}\sin ^{2}\theta \cos 2\varphi  \notag \\
&&+\frac{\partial _{\xi _{0}}C_{6}}{r^{6}}\sin ^{2}\theta \left\{ 1-\mathcal{%
F}_{\xi _{0}}^{2}(\varphi )+[1-\mathcal{F}_{\xi _{0}}(\varphi )]^{2}\cos
^{2}\theta \right\}  \notag \\
&&-2\frac{C_{6}}{r^{6}}\sin ^{2}\theta (\sin 4\xi _{0}\sin ^{2}\theta \cos
^{2}2\varphi +2\cos 2\xi _{0}\cos ^{2}\theta \cos 2\varphi ).
\end{eqnarray}
We use the relation%
\begin{eqnarray}
\sin ^{2}\theta \cos 2\varphi &=&2\sqrt{\frac{2\pi }{15}}(Y_{22}+Y_{2-2}),
\notag \\
\sin ^{2}\theta (1+\cos ^{2}\theta ) &=&\frac{4}{5}-\frac{8}{7}\sqrt{\frac{%
\pi }{5}}Y_{20}-\frac{16\sqrt{\pi }}{105}Y_{40},  \notag \\
\sin ^{2}\theta \cos ^{2}\theta \cos 2\varphi &=&\frac{2}{7}\sqrt{\frac{2\pi
}{15}}(Y_{22}+Y_{2-2})+\frac{4}{21}\sqrt{\frac{2\pi }{5}}(Y_{42}+Y_{4-2}),
\notag \\
\sin ^{4}\theta \cos ^{2}2\varphi &=&\frac{4}{15}-\frac{16}{21}\sqrt{\frac{%
\pi }{5}}Y_{20}+\frac{8\sqrt{\pi }}{105}Y_{40}+\frac{4}{3}\sqrt{\frac{2\pi }{%
35}}(Y_{44}+Y_{4-4})
\end{eqnarray}%
to re-express%
\begin{eqnarray}
\partial _{\xi _{0}}V_{\mathrm{eff}}(\mathbf{r}) &=&12\sqrt{\frac{2\pi }{15}}%
\frac{C_{3}}{r^{3}}\cos 2\xi _{0}(Y_{22}+Y_{2-2})+\frac{1}{r^{6}}\partial
_{\xi _{0}}C_{6}(\frac{4}{5}-\frac{8}{7}\sqrt{\frac{\pi }{5}}Y_{20}-\frac{16%
\sqrt{\pi }}{105}Y_{40})  \notag \\
&&-\frac{2}{r^{6}}(2C_{6}\cos 2\xi _{0}+\partial _{\xi _{0}}C_{6}\sin 2\xi
_{0})[\frac{2}{7}\sqrt{\frac{2\pi }{15}}(Y_{22}+Y_{2-2})+\frac{4}{21}\sqrt{%
\frac{2\pi }{5}}(Y_{42}+Y_{4-2})]  \notag \\
&&-\frac{1}{r^{6}}(2C_{6}\sin 4\xi _{0}+\partial _{\xi _{0}}C_{6}\sin
^{2}2\xi _{0})[\frac{4}{15}-\frac{16}{21}\sqrt{\frac{\pi }{5}}Y_{20}+\frac{8%
\sqrt{\pi }}{105}Y_{40}+\frac{4}{3}\sqrt{\frac{2\pi }{35}}(Y_{44}+Y_{4-4})]
\end{eqnarray}%
in terms of spherical harmonics.

In the eigen-basis $\{\left\vert \psi _{B}\right\rangle ,\left\vert \psi _{%
\mathbf{k}}\right\rangle \}$ of $H_{\mathrm{eff}}(\xi _{0})$, the
Hamiltonian becomes%
\begin{eqnarray}
H_{\mathrm{eff}}(\xi ) &=&\varepsilon _{B}\left\vert \psi _{B}\right\rangle
\left\langle \psi _{B}\right\vert +\int d\mathbf{k}\varepsilon _{\mathbf{k}%
}\left\vert \psi _{\mathbf{k}}\right\rangle \left\langle \psi _{\mathbf{k}%
}\right\vert +\delta \xi \sin \omega _{\mathrm{m}}t\int d\mathbf{k}%
\left\langle \psi _{\mathbf{k}}\right\vert \partial _{\xi _{0}}V_{\mathrm{eff%
}}(\mathbf{r})\left\vert \psi _{B}\right\rangle \left\vert \psi _{\mathbf{k}%
}\right\rangle \left\langle \psi _{B}\right\vert +\mathrm{H.c.}  \notag \\
&&+\delta \xi \sin \omega _{\mathrm{m}}t[\left\langle \psi _{B}\right\vert
\partial _{\xi _{0}}V_{\mathrm{eff}}(\mathbf{r})\left\vert \psi
_{B}\right\rangle \left\vert \psi _{B}\right\rangle \left\langle \psi
_{B}\right\vert +\int d\mathbf{k}d\mathbf{k}^{\prime }\left\langle \psi _{%
\mathbf{k}}\right\vert \partial _{\xi _{0}}V_{\mathrm{eff}}(\mathbf{r}%
)\left\vert \psi _{\mathbf{k}^{\prime }}\right\rangle \left\vert \psi _{%
\mathbf{k}}\right\rangle \left\langle \psi _{\mathbf{k}^{\prime
}}\right\vert ],
\end{eqnarray}%
where $\varepsilon _{\mathbf{k}}=k^{2}/M$.\ For the large dissociation
frequency $\omega _{\mathrm{m}}$, the rotating-wave-approximation (RWA)
results in%
\begin{equation}
H_{\mathrm{eff}}(\xi )\sim \varepsilon _{B}\left\vert \psi _{B}\right\rangle
\left\langle \psi _{B}\right\vert +\int d\mathbf{k}\varepsilon _{\mathbf{k}%
}\left\vert \psi _{\mathbf{k}}\right\rangle \left\langle \psi _{\mathbf{k}%
}\right\vert +e^{-i\omega _{\mathrm{m}}t}\int \frac{d\mathbf{k}}{(2\pi
)^{3/2}}g_{\mathbf{k}}\left\vert \psi _{\mathbf{k}}\right\rangle
\left\langle \psi _{B}\right\vert +\mathrm{H.c.},
\end{equation}%
where%
\begin{equation}
g_{\mathbf{k}}=i\frac{1}{2}\delta \xi (2\pi )^{3/2}\left\langle \psi _{%
\mathbf{k}}\right\vert \partial _{\xi _{0}}V_{\mathrm{eff}}\left\vert \psi
_{B}\right\rangle .
\end{equation}%
In the interaction picture, we obtain the dissociation Hamiltonian $H_{%
\mathrm{eff}}(\xi )\sim H_{\mathrm{diss}}$:%
\begin{equation}
H_{\mathrm{diss}}=(\varepsilon _{B}+\omega _{\mathrm{m}})\left\vert \psi
_{B}\right\rangle \left\langle \psi _{B}\right\vert +\int d\mathbf{k}%
\varepsilon _{\mathbf{k}}\left\vert \psi _{\mathbf{k}}\right\rangle
\left\langle \psi _{\mathbf{k}}\right\vert +\int \frac{d\mathbf{k}}{(2\pi
)^{3/2}}(g_{\mathbf{k}}\left\vert \psi _{\mathbf{k}}\right\rangle
\left\langle \psi _{B}\right\vert +\mathrm{H.c.})\mathrm{.}
\end{equation}

Using the numerical solutions%
\begin{eqnarray}
\psi _{B}(r) &=&\sum_{lm}\frac{\phi _{lm}(r)}{r}Y_{lm}(r),  \notag \\
\psi _{\mathbf{k}}(r) &=&\sum_{lm}\frac{\varphi _{\mathbf{k},lm}(r)}{r}%
Y_{lm}(r).
\end{eqnarray}%
of the bound-state and the scattering-state wavefunctions obtained from the
B-spline algorithm, we obtain the Frank-Condon factor%
\begin{eqnarray}
g_{\mathbf{k}} &=&i\frac{1}{2}\delta \xi (2\pi )^{3/2}\sum_{lml^{\prime
}m^{\prime }}\sqrt{\frac{2l^{\prime }+1}{2l+1}}\times  \notag \\
&&\{2\sqrt{6}C_{3}\cos 2\xi _{0}C_{20l^{\prime }0}^{l0}(C_{22l^{\prime
}m^{\prime }}^{lm}\delta _{mm^{\prime }+2}+C_{2-2l^{\prime }m^{\prime
}}^{lm}\delta _{mm^{\prime }-2})\int_{0}^{\infty }dr\frac{1}{r^{3}}\varphi _{%
\mathbf{k},lm}^{\ast }(r)\phi _{l^{\prime }m^{\prime }}(r)  \notag \\
&&+\partial _{\xi _{0}}C_{6}(\frac{4}{5}\delta _{ll^{\prime }}-\frac{4}{7}%
C_{20l^{\prime }0}^{l0}C_{20l^{\prime }m}^{lm}-\frac{8}{35}C_{40l^{\prime
}0}^{l0}C_{40l^{\prime }m}^{lm})\delta _{mm^{\prime }}\int_{0}^{\infty }dr%
\frac{1}{r^{6}}\varphi _{\mathbf{k},lm}^{\ast }(r)\phi _{l^{\prime
}m^{\prime }}(r)  \notag \\
&&-\frac{2}{7}(2C_{6}\cos 2\xi _{0}+\partial _{\xi _{0}}C_{6}\sin 2\xi _{0})[%
\sqrt{\frac{2}{3}}C_{20l^{\prime }0}^{l0}(C_{22l^{\prime }m^{\prime
}}^{lm}\delta _{mm^{\prime }+2}+C_{2-2l^{\prime }m^{\prime }}^{lm}\delta
_{mm^{\prime }-2})  \notag \\
&&+2\sqrt{\frac{2}{5}}C_{40l^{\prime }0}^{l0}(C_{42l^{\prime }m^{\prime
}}^{lm}\delta _{mm^{\prime }+2}+C_{4-2l^{\prime }m^{\prime }}^{lm}\delta
_{mm^{\prime }-2})]\int_{0}^{\infty }dr\frac{1}{r^{6}}\varphi _{\mathbf{k}%
,lm}^{\ast }(r)\phi _{l^{\prime }m^{\prime }}(r)  \notag \\
&&-(2C_{6}\sin 4\xi _{0}+\partial _{\xi _{0}}C_{6}\sin ^{2}2\xi _{0})[\frac{4%
}{15}\delta _{ll^{\prime }}\delta _{mm^{\prime }}-\frac{8}{21}C_{20l^{\prime
}0}^{l0}C_{20l^{\prime }m}^{lm}\delta _{mm^{\prime }}+\frac{4}{35}%
C_{40l^{\prime }0}^{l0}C_{40l^{\prime }m}^{lm}\delta _{mm^{\prime }}  \notag
\\
&&+\sqrt{\frac{8}{35}}C_{40l^{\prime }0}^{l0}(C_{44l^{\prime }m^{\prime
}}^{lm}\delta _{mm^{\prime }+4}+C_{4-4l^{\prime }m^{\prime }}^{lm}\delta
_{mm^{\prime }-4})]\int_{0}^{\infty }dr\frac{1}{r^{6}}\varphi _{\mathbf{k}%
,lm}^{\ast }(r)\phi _{l^{\prime }m^{\prime }}(r)\}.
\end{eqnarray}

\section{Dissociation spectrum and momentum distribution}

In this section, we derive the analytic formula for the dissociation
spectrum $P_{\mathrm{diss}}$ and the momentum distribution {$p_{\mathrm{m}}(%
\mathbf{k})$ of two dissociated molecules using the }dissociation
Hamiltonian $H_{\mathrm{diss}}$ for the small $\delta \xi $. We rewrite the
dissociation Hamiltonian as%
\begin{equation}
H_{\mathrm{diss}}=(\varepsilon _{B}+\omega _{\mathrm{m}})B^{\dagger }B+\int d%
\mathbf{k}\varepsilon _{\mathbf{k}}a_{\mathbf{k}}^{\dagger }a_{\mathbf{k}%
}+\int \frac{d\mathbf{k}}{(2\pi )^{3/2}}(g_{\mathbf{k}}a_{\mathbf{k}%
}^{\dagger }B+\mathrm{H.c.}),
\end{equation}%
which explicitly describes the spontaneous emission of the tetramer state to
the continuum. The dissociation spectrum $P_{\mathrm{diss}}=1-\left\vert
G_{B}(t)\right\vert ^{2}$ and the momentum distribution {$p_{\mathrm{m}}(%
\mathbf{k})$}$=\left\vert \int d\mathbf{p}\left\langle \mathbf{k}\left\vert
\psi _{\mathbf{p}}\right\rangle \right. G_{\mathbf{p}}(t)\right\vert ^{2}$
are determined by the time-ordered Green functions $G_{B}(t)=-i\left\langle
0\right\vert \mathcal{T}B(t)B^{\dagger }(0)\left\vert 0\right\rangle $ and $%
G_{\mathbf{p}}(t)=-i\left\langle 0\right\vert \mathcal{T}a_{\mathbf{p}%
}(t)B^{\dagger }(0)\left\vert 0\right\rangle $.

\begin{figure}[tbp]
\includegraphics[width=1.0\linewidth]{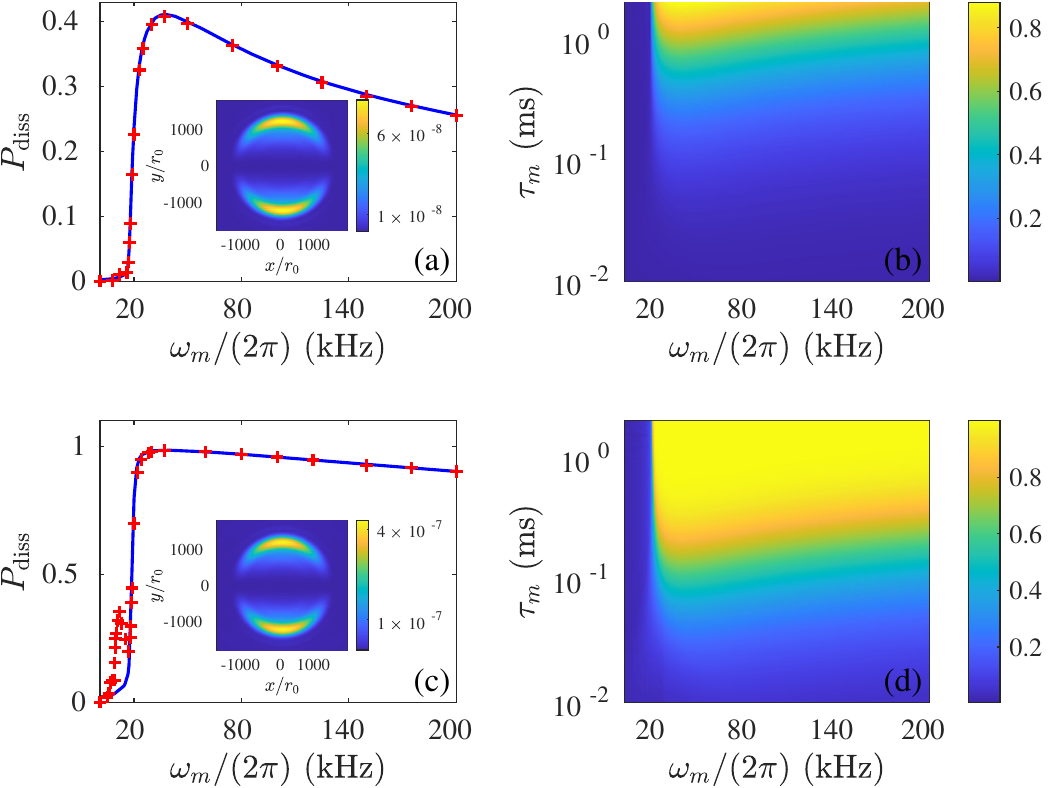}
\caption{Modulation dissociation spectra for the $p_y$-tetramer state $%
\left\vert \protect\psi_{B}^{(y)}\right\rangle$, where $(\Omega, \protect%
\delta)/(2\protect\pi)=(28.5, -9.5)\mathrm{MHz}$ and $\protect\xi_0=8^{\circ }$. The first and second rows displays $P_{ \mathrm{diss}}$ for $\protect\delta\protect\xi=0.5^{\circ}$ and $\protect\delta\protect\xi%
=1.4^{\circ}$, respectively. The first panel displays $P_{ \mathrm{diss}}$ at $\protect\tau_{\mathrm{m}}=0.5 \mathrm{ms}$, where the blue curves and red
crosses represent the RWA and exact results, respectively. The insets show
the coordinate distribution integrated along the $z$-direction at 3ms in TOF
after the dissociation with modulation frequency $\protect\omega_{\mathrm{m}%
}/(2\protect\pi)=37\mathrm{kHz}$. The second column displays $P_{\mathrm{diss%
}}$ under RWA in the $\protect\omega_{\mathrm{m}}$-$\protect\tau_{\mathrm{m}%
} $ plane.}
\label{pytetra}
\end{figure}

The Dyson expansion leads to the Green functions
\begin{eqnarray}
G_{B}(t) &=&\int \frac{d\omega }{2\pi }\frac{e^{-i\omega t}}{\omega
-\varepsilon _{0}-\Sigma (\omega )},  \notag \\
G_{\mathbf{p}}(t) &=&\frac{g_{\mathbf{p}}}{(2\pi )^{3/2}}\int \frac{d\omega
}{2\pi }\frac{e^{-i\omega t}}{\omega -\varepsilon _{\mathbf{p}}+i0^{+}}\frac{%
1}{\omega -\varepsilon _{0}-\Sigma (\omega )},
\end{eqnarray}%
where $\varepsilon _{0}=\varepsilon _{B}+\omega _{\mathrm{m}}$ and the
self-energy%
\begin{equation}
\Sigma (\omega )=\int \frac{d\mathbf{k}}{(2\pi )^{3}}\frac{\left\vert g_{%
\mathbf{k}}\right\vert ^{2}}{\omega -\varepsilon _{\mathbf{k}}+i0^{+}}.
\end{equation}

\begin{figure}[tbp]
\includegraphics[width=1.0\linewidth]{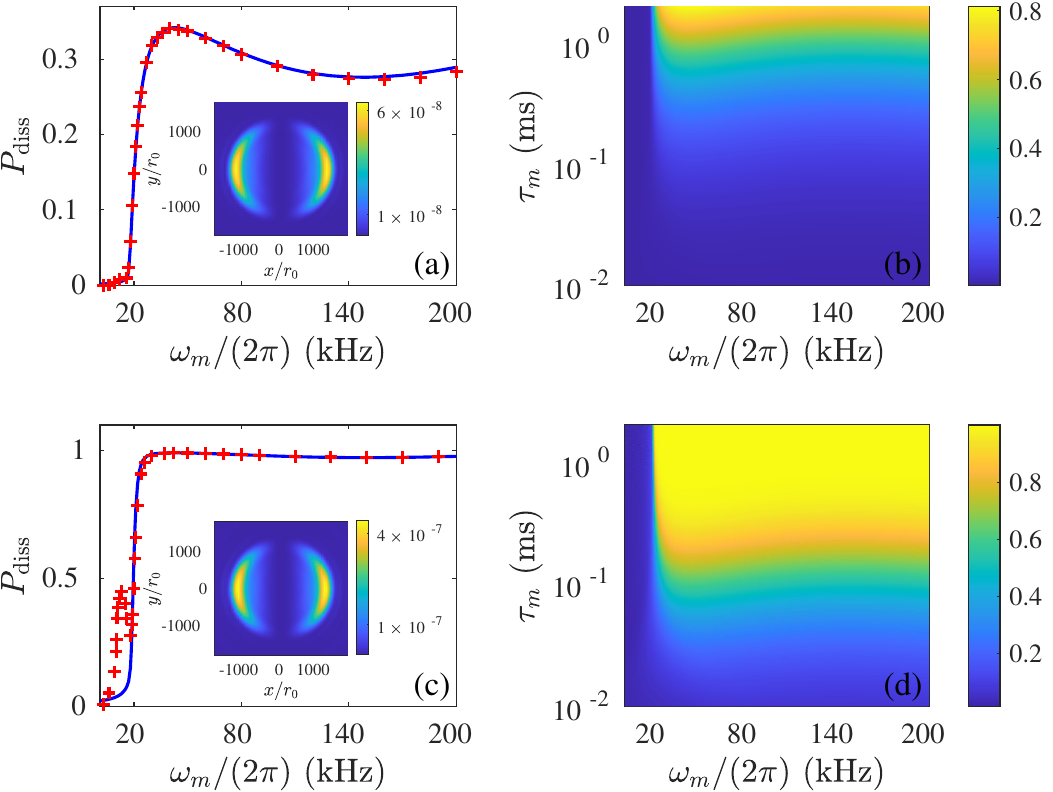}
\caption{Modulation dissociation spectra for the $p_x$-tetramer state $%
\left\vert \protect\psi_{B}^{(x)}\right\rangle$, where $(\Omega, \protect%
\delta)/(2\protect\pi)=(167.2, -9.5)\mathrm{MHz}$ and $\protect\xi_0=1.5^{\circ }$. The first and second rows displays $P_{ \mathrm{diss}}$ for $\protect\delta\protect\xi=0.3^{\circ}$ and $\protect\delta\protect\xi%
=1^{\circ}$, respectively. The first panel displays $P_{ \mathrm{diss}}$ at $\protect\tau_{\mathrm{m}}=0.5 \mathrm{ms}$, where the blue curves and red
crosses represent the RWA and exact results, respectively. The insets show
the coordinate distribution integrated along the $z$-direction at 3ms in TOF
after the dissociation with modulation frequency $\protect\omega_{\mathrm{m}%
}/(2\protect\pi)=37\mathrm{kHz}$. The second column displays $P_{\mathrm{diss%
}}$ under RWA in the $\protect\omega_{\mathrm{m}}$-$\protect\tau_{\mathrm{m}%
} $ plane.}
\label{pxtetra}
\end{figure}

Under the Markovian approximation, the self-energy reads%
\begin{equation}
\Sigma (\omega )=\int \frac{d\mathbf{k}}{(2\pi )^{3}}\mathrm{P}\frac{%
\left\vert g_{\mathbf{k}}\right\vert ^{2}}{\omega -\varepsilon _{\mathbf{k}}}%
-i\pi \int \frac{d\mathbf{k}}{(2\pi )^{3}}\left\vert g_{\mathbf{k}%
}\right\vert ^{2}\delta (\omega -\varepsilon _{\mathbf{k}})\sim -i\frac{1}{2}%
\gamma _{\mathrm{diss}},
\end{equation}%
where we omit the real part of $\Sigma (\omega )$ that renormalizes $%
\varepsilon _{0}$, and define the dissociation rate $\gamma _{\mathrm{diss}%
}=\int \frac{d\mathbf{k}}{(2\pi )^{2}}\left\vert g_{\mathbf{k}}\right\vert
^{2}\delta (\varepsilon _{0}-\varepsilon _{\mathbf{k}})$. It follows from
the residue theorem that%
\begin{eqnarray}
G_{B}(t) &=&-ie^{-\frac{1}{2}\gamma _{\mathrm{diss}}t}e^{-i\varepsilon
_{0}t},  \notag \\
G_{\mathbf{p}}(t) &=&-i\frac{g_{\mathbf{p}}}{(2\pi )^{3/2}}\frac{%
e^{-i\varepsilon _{\mathbf{p}}t}-e^{-i(\varepsilon _{0}-i\frac{1}{2}\gamma _{%
\mathrm{diss}})t}}{\varepsilon _{\mathbf{p}}-\varepsilon _{0}+i\frac{1}{2}%
\gamma _{\mathrm{diss}}},
\end{eqnarray}%
and%
\begin{eqnarray}
P_{\mathrm{diss}} &=&1-e^{-\gamma _{\mathrm{diss}}t},  \notag \\
{p_{\mathrm{m}}(\mathbf{k})} &=&\left\vert \int d\mathbf{p}\left\langle
\mathbf{k}\left\vert \psi _{\mathbf{p}}\right\rangle \right. \frac{g_{%
\mathbf{p}}}{(2\pi )^{3/2}}\frac{e^{-i\varepsilon _{\mathbf{p}%
}t}-e^{-i(\varepsilon _{0}-i\frac{1}{2}\gamma _{\mathrm{diss}})t}}{%
\varepsilon _{\mathbf{p}}-\varepsilon _{0}+i\frac{1}{2}\gamma _{\mathrm{diss}%
}}\right\vert ^{2}.
\end{eqnarray}%
In the limit $\gamma _{\mathrm{diss}}t_{F}\ll 1$, $P_{\mathrm{diss}}\sim
\gamma _{\mathrm{diss}}t$ agree with that from the Fermi Golden rule, and%
\begin{equation}
{p_{\mathrm{m}}(\mathbf{k})}\sim \frac{1}{4}\delta \xi ^{2}\left\vert
\left\langle \mathbf{k}\right\vert \partial _{\xi _{0}}V_{\mathrm{eff}%
}\left\vert \psi _{B}\right\rangle \right\vert ^{2}t^{2}.
\end{equation}

In Figs.~\ref{pytetra} and~\ref{pxtetra}, we compare the exact dissociation spectrum and that obtained using the effective Hamiltonian $H_{\mathrm{diss}}$ under RWA for the $p_y$ and $p_x$-tetramer states, respectively. For
small $\delta \xi$, the exact dissociation spectra via solving the time-dependent SE agrees with the RWA result quantitatively in the entire frequency domain. For large $\delta \xi$, the quantitative agreement between the exact result and the RWA result is also shown above the dissociation threshold, however, a below-threshold peak predicted by the exact result is displayed in $P_{\mathrm{diss}}$. 

The below-threshold peak corresponds to the multiple-photon process, which can be qualitatively described by the following effective model      
\begin{eqnarray}
H_{\mathrm{eff}} &=&(\varepsilon _{B}+\delta \omega _{B}\sin \omega _{%
\mathrm{m}}t)\left\vert \psi _{B}\right\rangle \left\langle \psi
_{B}\right\vert +\int d\mathbf{k}\varepsilon _{\mathbf{k}}\left\vert \psi _{%
\mathbf{k}}\right\rangle \left\langle \psi _{\mathbf{k}}\right\vert  \notag
\\
&&-2i\sin (\omega _{\mathrm{m}}t)\int \frac{d\mathbf{k}}{(2\pi )^{3/2}}g_{%
\mathbf{k}}\left\vert \psi _{\mathbf{k}}\right\rangle \left\langle \psi
_{B}\right\vert +\mathrm{H.c.,}
\end{eqnarray}%
where for small modulational frequency $\omega _{\mathrm{m}}$, the large modulation $\delta \xi$ induces the frequency fluctuation $\delta \omega _{B}=\delta \xi \left\langle \psi _{B}\right\vert
\partial _{\xi _{0}}V_{\mathrm{eff}}(\mathbf{r})\left\vert \psi
_{B}\right\rangle $ comparable to $\omega _{\mathrm{m}}$. Thus, the energy level fluctuation of the tetramer state is not negligable.

In the interacting picture, the Hamiltonian reads%
\begin{equation}
\bar{H}_{\mathrm{eff}}=\int d\mathbf{k}\varepsilon _{\mathbf{k}}\left\vert
\psi _{\mathbf{k}}\right\rangle \left\langle \psi _{\mathbf{k}}\right\vert
+e^{-i\varepsilon _{B}t}\sum\limits_{n,s=\pm 1}si^{n}J_{n}(\frac{\delta
\omega _{B}}{\omega _{\mathrm{m}}})e^{-i(n+s)\omega _{\mathrm{m}}t}\int
\frac{d\mathbf{k}}{(2\pi )^{3/2}}g_{\mathbf{k}}\left\vert \psi _{\mathbf{k}%
}\right\rangle \left\langle \psi _{B}\right\vert +\mathrm{H.c..}
\end{equation}%
When $\omega _{\mathrm{m}}$ is increasing from $0$ to the threshold value $%
\left\vert \varepsilon _{B}\right\vert $, the dominating process is the
transition from $\left\vert \psi _{B}\right\rangle $ to $\left\vert \psi _{%
\mathbf{k}}\right\rangle $ by absorbing $n_{0}+1$ modulation excitations.
Under the single frequency approximation, the Hamiltonian becomes%
\begin{equation}
\bar{H}_{\mathrm{eff}}\sim \int d\mathbf{k}\varepsilon _{\mathbf{k}%
}\left\vert \psi _{\mathbf{k}}\right\rangle \left\langle \psi _{\mathbf{k}%
}\right\vert +e^{-i[\varepsilon _{B}+(n_{0}+1)\omega _{\mathrm{m}%
}]t}i^{n_{0}}J_{n_{0}}(\frac{\delta \omega _{B}}{\omega _{\mathrm{m}}})\int
\frac{d\mathbf{k}}{(2\pi )^{3/2}}g_{\mathbf{k}}\left\vert \psi _{\mathbf{k}%
}\right\rangle \left\langle \psi _{B}\right\vert +\mathrm{H.c..}
\end{equation}%
By eliminating the time-dependent phase, we obtain%
\begin{eqnarray}
\bar{H}_{\mathrm{eff}} &\sim &[\varepsilon _{B}+(n_{0}+1)\omega _{\mathrm{m}%
}]\left\vert \psi _{B}\right\rangle \left\langle \psi _{B}\right\vert +\int d%
\mathbf{k}\varepsilon _{\mathbf{k}}\left\vert \psi _{\mathbf{k}%
}\right\rangle \left\langle \psi _{\mathbf{k}}\right\vert  \notag \\
&&+i^{n_{0}}J_{n_{0}}(\frac{\delta \omega _{B}}{\omega _{\mathrm{m}}})\int
\frac{d\mathbf{k}}{(2\pi )^{3/2}}g_{\mathbf{k}}\left\vert \psi _{\mathbf{k}%
}\right\rangle \left\langle \psi _{B}\right\vert +\mathrm{H.c.,}
\end{eqnarray}%
which has the same form as $H_{\mathrm{diss}}$, where the single excitation
frequency $\omega _{\mathrm{m}}$ is replaced by the multi-excitation
frequency $(n_{0}+1)\omega _{\mathrm{m}}$ and $g_{\mathbf{k}}\rightarrow
i^{n_{0}}J_{n_{0}}(\frac{\delta \omega _{B}}{\omega _{\mathrm{m}}})g_{%
\mathbf{k}}$. The dissociation studied in Fig. 4b of the main text is a two-photon process, where $n_{0}=1$. 

\end{document}